\documentclass[pre,aps,10pt,twocolumn,floatfix,superscriptaddress]{revtex4-1}

\usepackage[utf8]{inputenc}
\usepackage[nice]{nicefrac}
\usepackage{graphicx,color}
\usepackage{amsmath,amssymb,bm,bbm}
\usepackage{amsmath}
\usepackage{braket}
\usepackage{xcolor}
\usepackage{float}

\usepackage[nice]{nicefrac}

\usepackage{physics} 

\usepackage[T1]{fontenc}
\usepackage[utf8]{inputenc}

\usepackage[plainpages=false,pdfpagelabels,colorlinks=true,linkcolor=red,urlcolor=blue,citecolor=blue,pdftitle={},pdfauthor={},pdfdisplaydoctitle=true,pdfduplex=DuplexFlipLongEdge]{hyperref}

\definecolor{max}{HTML}{03A678}
\definecolor{rafal}{HTML}{161226}

\definecolor{darkgreen}{rgb}{0,0.60,.2}
\definecolor{darkblue}{rgb}{0.1,0.3,1}

\usepackage{physics} 
\usepackage[normalem]{ulem} 
\usepackage[T1]{fontenc}
\usepackage[utf8]{inputenc}

\usepackage{amsthm}
\usepackage{amssymb}

\date{\today}

\begin{document}

\title{Average entanglement entropy of midspectrum eigenstates of\\ quantum-chaotic interacting Hamiltonians}

\author{M. Kliczkowski}
\affiliation{Institute of Theoretical Physics, Faculty of Fundamental Problems of Technology, Wrocław University of Science and Technology, 50-370 Wrocław, Poland\looseness=-3}
\affiliation{Department of Physics, The Pennsylvania State University, University Park, Pennsylvania 16802, USA}
\author{R. Świętek}
\affiliation{Department of Theoretical Physics, J. Stefan Institute, SI-1000 Ljubljana, Slovenia\looseness=-1}
\affiliation{Department of Physics, Faculty of Mathematics and Physics, University of Ljubljana, SI-1000 Ljubljana, Slovenia\looseness=-1}
\author{L. Vidmar}
\affiliation{Department of Theoretical Physics, J. Stefan Institute, SI-1000 Ljubljana, Slovenia\looseness=-1}
\affiliation{Department of Physics, Faculty of Mathematics and Physics, University of Ljubljana, SI-1000 Ljubljana, Slovenia\looseness=-1}
\author{M. Rigol}
\affiliation{Department of Physics, The Pennsylvania State University, University Park, Pennsylvania 16802, USA}

\begin{abstract}
To which degree the average entanglement entropy of midspectrum eigenstates of quantum-chaotic interacting Hamiltonians agrees with that of random pure states is a question that has attracted considerable attention in the recent years. While there is substantial evidence that the leading (volume-law) terms are identical, which and how subleading terms differ between them is less clear. Here we carry out state-of-the-art full exact diagonalization calculations of clean spin-1/2 XYZ and XXZ chains with integrability breaking terms to address this question in the absence and presence of $U(1)$ symmetry, respectively. We first introduce the notion of maximally chaotic regime, for the chain sizes amenable to full exact diagonalization calculations, as the regime in Hamiltonian parameters in which the level spacing ratio, the distribution of eigenstate coefficients, and the entanglement entropy are closest to the random matrix theory predictions. In this regime, we carry out a finite-size scaling analysis of the subleading terms of the average entanglement entropy of midspectrum eigenstates when different fractions $\nu$ of the spectrum are included in the average. We find indications that, for $\nu\rightarrow0$, the magnitude of the negative $O(1)$ correction is only slightly greater than the one predicted for random pure states. For finite $\nu$, following a phenomenological approach, we derive a simple expression that describes the numerically observed $\nu$ dependence of the $O(1)$ deviation from the prediction for random pure states.
\end{abstract}

\maketitle

\section{Introduction} \label{sec:introduction}

Entanglement is one of the most fundamental and intriguing features of quantum mechanics~\cite{bell1964einstein, horodecki_09}. Since the early 2000s, we have learned that in physical Hamiltonians there are qualitative differences between entanglement in ground states, which typically exhibit an ``area-law'' entanglement entropy~\cite{eisert2010colloquium}, and in highly excited energy eigenstates, which typically exhibit a ``volume-law'' entanglement entropy~\cite{bianchi_hackl_22}. Entanglement has also been conjectured to serve as a diagnostic for quantum chaos and integrability~\cite{leblond_mallayya_19}. Using full exact diagonalization calculations, the average bipartite entanglement entropies of highly excited eigenstates of spin-1/2 XXZ chains were shown to behave qualitatively differently at and away from integrability~\cite{leblond_mallayya_19}. Specifically, while the average in both regimes exhibits a leading volume-law term, the coefficient of the volume in that term was found to be maximal for midspectrum eigenstates in the quantum-chaotic regime (consistent with findings in earlier works~\cite{vidmar_rigol_17, garrison_grover_18}; see also Refs.~\cite{beugeling_andreanov_15, yang_chamon_15, dymarsky_lashkari_18}) and lower than maximal and subsystem-fraction dependent at integrability (as found for quadratic models and integrable models mappable onto quadratic ones~\cite{vidmar_hackl_17, vidmar_hackl_18, liu_chen_18, hackl_vidmar_19, jafarizadeh_rajabpour_19, lydzba_rigol_20, lydzba_rigol_21, bianchi_hackl_21}).

Following on the previously mentioned studies, our focus in this work is the entanglement entropy of highly excited eigenstates of clean spin-1/2 quantum-chaotic interacting Hamiltonians. We consider chains with $L$ sites (we change $L$ to carry out scaling analyses) with periodic and open boundary conditions. For pure quantum states $|\psi\rangle$, which we will take to be Hamiltonian eigenstates, we study the (bipartite) entanglement entropy of a subsystem $A$ (with $L_A$ contiguous sites) after tracing out the complement $B$ (with $L_B=L-L_A$ contiguous sites), resulting in 
\begin{equation}
\hat \rho_A=\mathrm{Tr}_{B}|\psi\rangle\langle\psi|.
\end{equation}
The von Neumann entanglement entropy (in short, the entanglement entropy) of subsystem $A$ is 
\begin{equation}\label{Neumann.entropy}
    S_A=-\mathrm{Tr}(\hat \rho_A\ln\hat \rho_A).
\end{equation}
The entanglement entropy of random pure states with the same Hilbert space in a lattice with $L$ sites (whose spacial configuration is irrelevant) is a natural counterpart to compare to the entanglement entropy of eigenstates of quantum-chaotic interacting Hamiltonians~\cite{bianchi_hackl_22}. By now, there is strong evidence that the leading term in the average entanglement entropy is the same for midspectrum eigenstates and for random pure states~\cite{bianchi_hackl_22}. An important question is then to which degree the average entanglement entropy of those midspectrum eigenstates agrees, beyond the leading volume-law term, with the average over random pure states. 

For spin-1/2 chains with $U(1)$ symmetry (particle-number conservation in the spinless fermions language) this question was explored by two of us (L.V. and M.R.) in Ref.~\cite{vidmar_rigol_17}. Away from the zero magnetization sector (``half-filling'' for fermions) the main finding was that, when one traces out 1/2 of the lattice sites, the average entanglement entropy of random pure states exhibits a first subleading term that scales with the square root of the number of sites in the lattice. The numerical calculations then indicated that, remarkably, the average entanglement entropy of the midspectrum eigenstates of the quantum-chaotic spin-1/2 chain exhibit the same subleading term. At zero magnetization, when one traces out 1/2 of the lattice sites, the first subleading term in the average entanglement entropy of random pure states is $O(1)$~\cite{bianchi_hackl_22}. The numerical calculations in Ref.~\cite{vidmar_rigol_17} indicated that the same is true about the average entanglement entropy of the midspectrum eigenstates, with a value of the constant that is close to that for random pure states. 

The leading terms of the average entanglement entropy of random pure states with fixed total magnetization $S^z$, magnetization per site $s^z=S^z/L$, or, equivalently, at spinless fermions filling $n=s^z+1/2$, were fully derived in Ref.~\cite{bianchi_hackl_22} using methods introduced in Ref.~\cite{bianchi_dona_19}:
\begin{align}\label{eq:nleading}
	\langle S_A\rangle_{n}=&-[n\ln n+(1-n)\ln(1-n)]\, L_A\nonumber\\
	&-\sqrt{\frac{n(1-n)}{2\pi}}\left|\ln\left(\frac{1-n}{n}\right)\right|\delta_{f,\frac{1}{2}}\sqrt{L}\nonumber\\
	&+\frac{f+\ln(1-f)}{2}-\frac{1}{2}\delta_{f,\frac{1}{2}}\delta_{n,\frac{1}{2}}+o(1),
\end{align}
where $L_A\leq \frac{L}{2}$, and $o(1)$ is used for terms that vanish in the thermodynamic limit. In Eq.~\eqref{eq:nleading}, $f=L_A/L$ stands for the ``subsystem fraction.'' To obtain the results for $L_A>\frac{L}{2}$ ($f>\frac{1}{2}$), one just needs to replace $L_A\rightarrow L-L_A$ in Eq.~\eqref{eq:nleading}.

It is important to remark that because of the $U(1)$ symmetry, in Eq.~\eqref{eq:nleading} there is an $O(1)$ ``mean-field'' correction to the average entanglement entropy of random pure states at all values of the magnetization and subsystem fractions, which was derived in Ref.~\cite{vidmar_rigol_17},
\begin{equation}\label{eq:mfo1}
S_\text{MF}^{(1)}=\frac{f+\ln(1-f)}{2}.
\end{equation}
The fact that this is the only $O(1)$ correction to the leading volume-law term away from $f=1/2$ was proved later in Ref.~\cite{bianchi_hackl_22}. At $f=1/2$, and {\it only} at zero magnetization, the additional $-1/2$ correction is the same as that found by Page in the absence of $U(1)$ symmetry~\cite{page_1993b}. In the latter case, the average entanglement entropy over random states reads
\begin{equation}\label{eq:page_inf}
\langle S_A\rangle=L_A \ln 2  - \frac{1}{2} \delta_{f,\frac{1}{2}} + o(1).
\end{equation}

Recent works have attempted to identify the reasons behind, and quantify the differences, between the average entanglement entropy over random pure states and over Hamiltonian eigenstates in the absence of $U(1)$ symmetry~\cite{huang_19, Haque_Khaymovich_2022, huang_21, huang_22}. In Ref.~\cite{huang_19}, Huang conjectured (and provided some numerical evidence) that the average entanglement entropy over all eigenstates of local quantum-chaotic interacting Hamiltonians has the form
\begin{equation}\label{eq:w=0}
\bar{\text{S}}_A=L_A \ln{2}+ \frac{\ln(1-f)}{2}-\frac{2}{\pi}\delta_{f,1/2}.
\end{equation}
This formula was derived under an assumption of chaoticity and locality of the Hamiltonian. The result in Eq.~\eqref{eq:w=0} coincides with the one obtained in Ref.~\cite{bianchi_hackl_22} for the average entanglement entropy of random pure states in the presence of $U(1)$ symmetry when all (properly weighted) magnetization sectors are included in the average.

Following on Ref.~\cite{huang_19}, and on numerical results reported in Ref.~\cite{Haque_Khaymovich_2022}, Huang conjectured (and provided some numerical evidence) that the average entanglement entropy over midspectrum eigenstates of (local) quantum-chaotic interacting Hamiltonians at $f=1/2$ has the form~\cite{huang_21}
\begin{eqnarray}\label{eq:nu}
\bar{\text{S}}_A^{(\nu)}&=&\frac{L-1}{2}\ln2+\frac{2\left[e^{-(\erf^{-1}\nu)^2}-1\right]}{\nu\pi}\nonumber \\&&+\frac{\left[e^{-(\erf^{-1}\nu)^2}+2\nu-2\right]\erf^{-1}\nu}{2\nu\sqrt\pi},
\end{eqnarray}
where $\nu$ is the fraction of the Hilbert space over which one carries out the average. For $\nu=1$, one recovers Eq.~\eqref{eq:w=0}, while for $\nu=0^+$ one obtains
\begin{equation}\label{eq:nu0}
	\bar{\text{S}}_A^{(0^+)} =\frac{L}{2}\ln 2 -\frac{\ln2}{2}-\frac{1}{4},
\end{equation}
which coincides with the result in Eq.~\eqref{eq:nleading} at $f=1/2$ and $n=1/2$, i.e.,  Eq.~\eqref{eq:nu0} is identical to the result for the average over random pure states with fixed magnetization $s^z=0$ at $f=1/2$. Building on these findings, in Ref.~\cite{nieva2023} it was argued that at $\nu=0^+$ energy conservation in quantum-chaotic local Hamiltonians~\cite{murthy_srednicki_19a} plays a similar role to that of $U(1)$ symmetry in random pure states.

In this work, we study the subleading corrections to the average entanglement entropy of midspectrum eigenstates in quantum-chaotic interacting spin-1/2 chains without $U(1)$ symmetry~\cite{huang_19, Haque_Khaymovich_2022, huang_21, huang_22, nieva2023} and with $U(1)$ symmetry~\cite{vidmar_rigol_17}. We carry out state-of-the-art numerical calculations of clean spin-1/2 XYZ and XXZ models with integrability breaking terms in chains with periodic and open boundary conditions. Using various quantum chaos indicators, we first scan a wide range of parameters for those models and introduce the concept of the maximally chaotic regime. Namely, a regime in Hamiltonian parameters in which, for the chain sizes that are accessible to full exact diagonalization calculations, the quantum-chaos indicators considered are closest to the random matrix theory predictions. It is in this regime that we find that the midspectrum energy eigenstates exhibit the greatest entanglement entropy. We then carry out finite-size scaling analyzes of the average eigenstate entanglement entropy in the maximally chaotic regime.

The paper is organized as follows. In Sec.~\ref{sec:model}, we introduce the models under consideration and discuss details about our numerical calculations. The maximally chaotic regime is identified in Sec.~\ref{sec:chaos}. The results for the finite-size scaling analyzes of the average entanglement entropy of midspectrum energy eigenstates are presented in Sec.~\ref{sec:entro}. We summarize our results in Sec.~\ref{sec:conclusions}.

\section{Models} \label{sec:model}

We study the clean spin-$1/2$ XYZ chain ($\hat H_{\rm XYZ}$) with nearest- ($\hat H_1$) and next-nearest- ($\hat H_2$) neighbors interactions in a magnetic field ($\hat H_F$):
\begin{eqnarray}\label{model:eq1}
        &&\hspace{-0.4cm}\hat H_{\rm XYZ}= \hat H_1+\hat H_2 + \hat H_F, \\
        &&\hspace{-0.4cm}\hat H_1 = J_1\sum_{\ell}[(1-\eta) \hat s_\ell^x \hat s_{\ell+1}^x+(1+\eta) \hat s_\ell^y \hat s_{\ell+1}^y + \Delta_1 \hat s_\ell^z \hat s_{\ell+1}^z],\nonumber \\
        &&\hspace{-0.4cm}\hat H_2 = J_2\sum_{\ell}[(1-\eta) \hat s_\ell^x \hat s_{\ell+2}^x+(1+\eta) \hat s_\ell^y \hat s_{\ell+2}^y + \Delta_2 \hat s_\ell^z \hat s_{\ell+2}^z],\nonumber \\
        &&\hspace{-0.4cm}\hat H_F = \sum_{\ell} (h^z \hat s^z_\ell + h^x \hat s^x_\ell)\nonumber,
\end{eqnarray}
where $\hat s_{\ell}^{\tau}$, with $\tau=x,y,z$, are the spin-1/2 operators at site $\ell$. We fix the five parameters $J_1=1$ (unit of energy), $\eta = 0.5$, $\Delta_1=0.3$, $\Delta_2 = 0.3$, and $h^x=0.3$, whereas the remaining two parameters $J_2,\,h_z$ are determined by the analysis discussed in Sec.~\ref{sec:chaos}. 

In our full exact diagonalization calculations, we consider chains with periodic boundary conditions, i.e., $\hat s_{L+1}^{\tau} = \hat s_1^\tau$ and $\hat s_{L+2}^{\tau} = \hat s_2^\tau$ ($\tau=x,y,z$), and with open boundary conditions. With periodic boundary conditions, the Hamiltonian in Eq.~\eqref{model:eq1} exhibits translation symmetry. We resolve it resulting in a block-diagonal structure of the Hamiltonian in which each block is labeled by the total quasimomentum $k$ and has a total number of states $\sim 2^L/L$. The blocks with $k=0$ and $\pi$ are further split using the reflection symmetry also present in the Hamiltonian, resulting in sub-blocks with $\sim 2^L/(2L)$ states. For open boundary conditions, only reflection symmetry is present, resulting in a splitting of the Hamiltonian in two blocks with $\sim 2^L/2$ states. For periodic boundary conditions, the largest chains that we study have $L=22$ for $k=0$ and $\pi$, and $L=20$ for all the other total quasimomentum sectors. For open boundary conditions, the largest chains that we study have $L=18$.

The second model we study is the clean spin-$1/2$ XXZ chain ($\hat H_{\rm XXZ}$) with nearest- ($\hat H'_1$) and next-nearest- ($\hat H'_2$) neighbors interactions:
\begin{eqnarray}\label{model:eq2}
        &&\hat H_{\rm XXZ}= \hat H'_1+\hat H'_2,\\
        &&\hat H'_1 = J_1\sum_{\ell}[\hat s_\ell^x \hat s_{\ell+1}^x + \hat s_\ell^y \hat s_{\ell+1}^y + \Delta_1 \hat s_\ell^z \hat s_{\ell+1}^z],\nonumber \\
        &&\hat H'_2 = J_2\sum_{\ell}[\hat s_\ell^x \hat s_{\ell+2}^x+ \hat s_\ell^y \hat s_{\ell+2}^y + \Delta_2 \hat s_\ell^z \hat s_{\ell+2}^z], \nonumber 
\end{eqnarray}
which is obtained from Eq.~\eqref{model:eq1} by setting $\eta=h^x=h^z=0$. We fix $J_1=1$ and $\Delta_2 = 0.3$, whereas the remaining parameters $J_2,\, \Delta_1$ are determined by the analysis discussed in Sec.~\ref{sec:chaos}. 

The spin-$1/2$ XXZ chain has an additional $U(1)$ symmetry, so that the total magnetization ($\hat S^z=\sum_{\ell}\hat s_\ell^z$) is a conserved quantity. At zero magnetization, it further has a ($Z_2$) spin-flip symmetry. We resolve both symmetries in our calculations, which are carried out in the zero total magnetization sector. For chains with periodic boundary conditions, this further reduces the number of states in the blocks that need to be diagonalized to $\sim\binom{L}{L/2}/(4L)$ at $k=0$ and $\pi$, and $\sim\binom{L}{L/2}/(2L)$ for all other total quasimomentum sectors. For the chains with open boundary conditions, we need to fully diagonalize blocks with $\sim\binom{L}{L/2}/4$ states. For periodic boundary conditions, we study chains with up to $L=26$ for $k=0$ and $\pi$, and up to $L=24$ for all other $k$ sectors. For open boundary conditions, we study chains with up to $L=20$.

Unless stated otherwise, the exact diagonalization results reported correspond to averages over all symmetry blocks for any given chain size $L$, and the number of eigenstates reported in the context of the averages is the one taken from each symmetry block.

\section{Maximally chaotic regime} \label{sec:chaos}

Several quantities, associated to the eigenenergies or to the energy eigenstates, have been traditionally used to quantify ``quantum chaos'' in many-body interacting Hamiltonians~\cite{dalessio_kafri_16}. They are computed in model Hamiltonians and compared to the predictions from random matrix theory (RMT). Their agreement, or the improvement of their agreement with increasing system size, are considered a hallmark of many-body quantum chaos. 

Since various limits of one-dimensional chains (such as the ones considered here) are integrable, when carrying out scaling analyses to make predictions about generic quantum-chaotic interacting models it is desirable to be as ``far away'' as possible from integrable points. In this spirit, in this section we identify the {\it maximally chaotic regime} for the chain sizes that we can study using full exact diagonalization calculations of the Hamiltonians in Sec.~\ref{sec:model}. The maximally chaotic regime is the regime in the model parameters in which we find the closest agreement between the exact diagonalization results and the RMT predictions.

\subsection{Level spacing ratio} \label{sec:levelspacing}

The statistical properties of the eigenenergies  $\{E_\alpha\}$ of quantum many-body Hamiltonians are one of the most commonly used indicators of quantum chaos~\cite{dalessio_kafri_16}. One of the simplest and most studied associated quantity is the distribution of the ratios of consecutive level spacings, defined as~\cite{oganesyan_huse_07}:
\begin{equation}\label{chaos:eq1}
    r_\alpha=\frac{\min\{\delta_\alpha,\delta_{\alpha+1}\}}{\max\{\delta_\alpha,\delta_{\alpha+1}\}} \;,
\end{equation}
where $\delta_\alpha = E_{\alpha} - E_{\alpha-1}$ is the energy difference between consecutive levels. The RMT prediction for the average of $r_\alpha$ in the Gaussian orthogonal ensemble (GOE) is $\bar r_{\rm GOE} = 0.5307$~\cite{atas_bogomolny_13}, and this is the result one expects to obtain in quantum-chaotic interacting Hamiltonians.

\begin{figure}[!t]
\includegraphics[width=0.98\columnwidth]{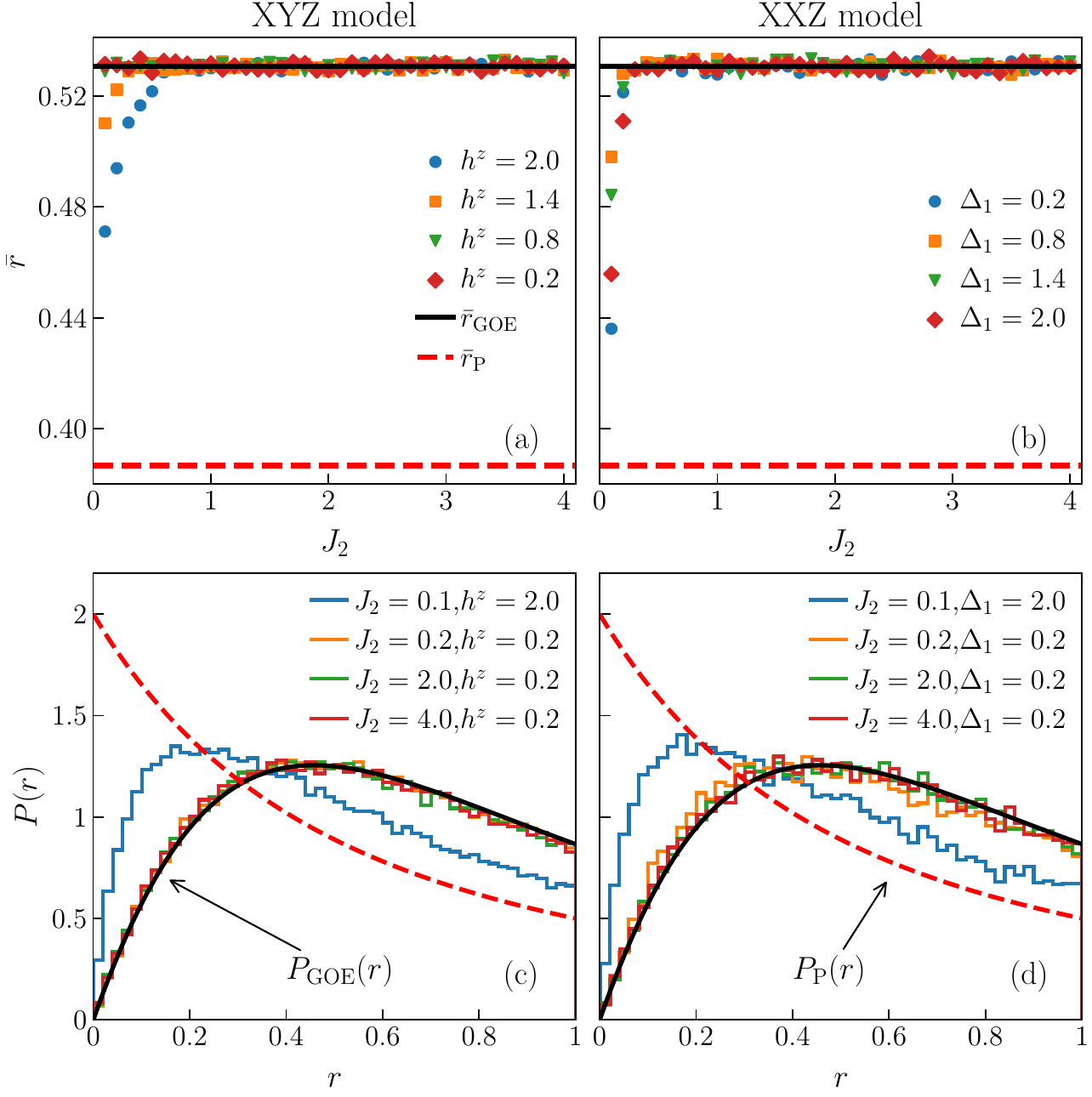}
\caption{Statistics of the level spacing ratio, see Eq.~\eqref{chaos:eq1}. [(a) and (b)] Average level spacing ratio $\bar r$ for $50\%$ of midspectrum eigenstates in the XYZ ($L=18$) and XXZ ($L=20$) models, respectively. The solid (dashed) horizontal line denotes the GOE (Poisson distribution) prediction $\bar r_{\rm GOE} = 0.5307$ ($\bar r_{\rm P} = 0.3867$). [(c) and (d)] Corresponding distributions $P(r)$ for four sets of Hamiltonian parameters. The solid [dashed] line denotes the GOE [Poisson distribution] prediction $P_{\rm GOE}(r) = \frac{27}{4}(r+r^2)/(1+r+r^2)^{5/2}$ [$P_{\rm P}(r) = 2/(1+r)^2$].}
\label{fig:levelspac}
\end{figure}

In Fig.~\ref{fig:levelspac}(a) [Fig.~\ref{fig:levelspac}(b)], we show the average level spacing ratio $\bar r$ for the XYZ [XXZ] model as a function of $J_2$ for different values of $h_z$ [$\Delta_1$] in a chain with $L=18$ [$L=20$]. For both the XYZ and XXZ models, the deviations from the GOE predictions are greatest for small values of $J_2$. For the XYZ chain, we also find that large values of $h^z$ extend the regime in $J_2$ in which greater deviations are seen from the RMT prediction, while in the XXZ model the extent of that region is quite insensitive to the value of $\Delta_1$ for the range of parameters shown. In Figs.~\ref{fig:levelspac}(c) and~\ref{fig:levelspac}(d), we show exemplary distributions of $r$ for four sets of Hamiltonian parameters for the XYZ and XXZ models, respectively, and how they compare to the predictions for the GOE and the Poisson distribution. 

Our results in Fig.~\ref{fig:levelspac} show that there is a broad regime (of the two Hamiltonian parameters that we have not fixed in both models) in which there is a nearly perfect agreement with the GOE predictions (up to the statistical fluctuations necessarily present in our finite samples of eigenenergies). The parameters of both models that are fixed in Fig.~\ref{fig:levelspac} were selected to ensure such an agreement with the GOE predictions for a wide range of the parameters left to fix. We spare the readers a discussion of that tedious broader exploration.

\subsection{Eigenstate coefficients} \label{sec:coeff}

Next, we study the distribution of eigenstate coefficients, which we find to be a more sensitive diagnostic of quantum chaos than $r$. Specifically, given a Hamiltonian eigenstate $\psi_\alpha=\sum_m c^\alpha_m\ket{m}$, expanded in the computational basis \{$\ket{m}$\}, we study the distribution of $c^\alpha_m$ over the midspectrum eigenstates. 

Due to the presence of translation symmetry, the coefficients are complex in all total quasimomentum sectors except for $k=0$ and $\pi$. Hence, we proceed in two steps. In the first step, we consider eigenstates from a single symmetry block. If the eigenstate coefficients are complex, then we rescale both the real and the imaginary parts separately by their corresponding standard deviations $\sigma$ [i.e., $\Re(c_m^\alpha) \to \Re(c_m^\alpha)/\sigma$ and $\Im(c_m^\alpha) \to \Im(c_m^\alpha)/\sigma$, where $\sigma \propto 1/\sqrt{2D}$ with $D$ the number of states in the corresponding sector], while in case of real coefficients we directly rescale the coefficients (i.e., $c_m^\alpha \to c_m^\alpha/\sigma$, where $\sigma \propto 1/\sqrt{D}$ with $D$ the number of states in the corresponding sector). In the second step, we collect the rescaled real and imaginary parts of all the coefficients and study the statistics of their absolute values, denoted as $z$ in this paper. The probability density function (PDF) of $z$, $P(z)$, is contrasted to a Gaussian PDF of the form
\begin{equation}\label{coeff:eq1}
    \bar{P}(z)=\frac{2}{\sqrt{2\pi}}e^{-\frac{z^2}{2}} \;,
\end{equation}
where a prefactor of 2 appears in the numerator because we study the distribution of absolute values, see Appendix~\ref{app::dist}. We split and analyze the  distributions of the coefficients this way to avoid having to deal with different distributions for the real and complex sectors. The distribution of the absolute values of coefficients $c_m^\alpha$ in the complex sectors follows a chi-squared distribution, see Appendix~\ref{app::dist} for details.

\begin{figure}[!t]
\includegraphics[width=0.98\columnwidth]{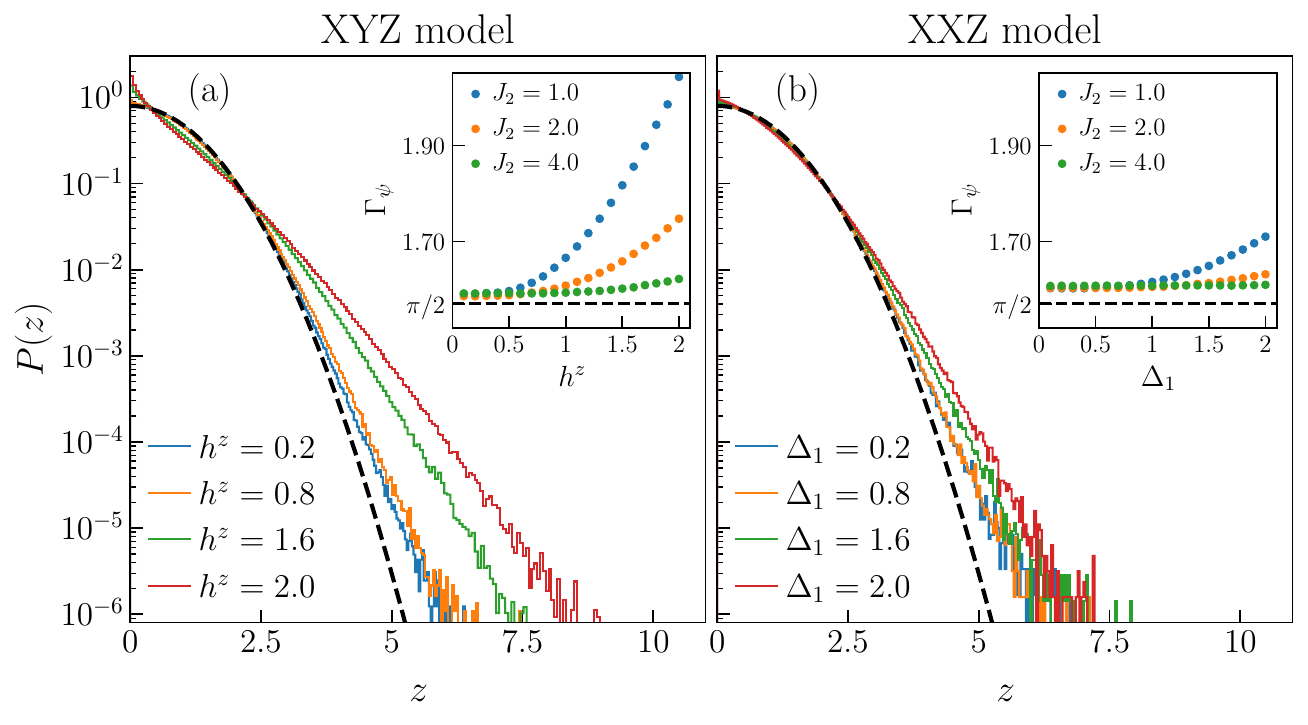}
\caption{Distribution of eigenstate coefficients $z$ (see text) obtained using $100$ midspectrum eigenstates from each total quasimomentum sector. In our calculations we take $J_2=1.0$, for (a) the XYZ model ($L=18$) and (b) the XXZ model ($L=20$). The dashed lines show $\bar P(z)$ from Eq.~\eqref{coeff:eq1}. Insets: $\Gamma_\psi$ from Eq.~\eqref{coeff:eq2} for distributions such as the ones shown in the main panels for three values of $J_2$ ($J_2=1.0$, $2.0$, and $4.0$). The results for $\Gamma_\psi$ in the insets are plotted as functions of (a) $h^z$, and (b) $\Delta_1$.}
\label{fig:coefficients}
\end{figure}

The main panels of Fig.~\ref{fig:coefficients} show the PDFs $P(z)$ for the XYZ [Fig.~\ref{fig:coefficients}(a)] and XXZ [Fig.~\ref{fig:coefficients}(b)] models. We take $J_2=1.0$, for which the average level spacing ratio $\bar r$ in Fig.~\ref{fig:levelspac} matches the GOE prediction for the model parameters under investigation. The results in Fig.~\ref{fig:coefficients} show that the distribution of eigenstate coefficients is a more sensitive probe of many-body quantum chaos than $r$. Specifically, all the PDFs in Fig.~\ref{fig:coefficients} show deviations from the Gaussian distribution $\bar P(z)$, which are most visible in the tails of the distributions. Several works~\cite{luitz_barlev_16, wang2018, beugeling_baecker_18, baecker_haque_19, Haque_Khaymovich_2022} have studied distributions of eigenstate coefficients and also observed deviations from the Gaussian distribution.

We find that, for $J_2=1.0$, the PDFs $P(z)$ are closest to the Gaussian function $\bar P(z)$ at small $h^z<1$ in the XYZ model and at small $\Delta_1<1$ in the XXZ model. In order to quantify the deviations from the Gaussian PDF, as done in Ref.~\cite{leblond_mallayya_19}, we compute the ratio
\begin{equation}\label{coeff:eq2}
    \Gamma_\psi=\frac{\expval{z^2}}{\expval{z}^2} \;,
\end{equation}
which yields $\Gamma_\psi = \pi/2$ for the Gaussian PDF from Eq.~\eqref{coeff:eq1}. Numerical results for $\Gamma_\psi$ are shown in the insets of Fig.~\ref{fig:coefficients} for $J_2=1.0$, 2.0, and 4.0 as functions of (a) $h^z$ for the XYZ model and (b) $\Delta_1$ for the XXZ model. From those plots we conclude that, for the closest agreement with the RMT prediction, one needs small values of $h^z$ for the XYZ model and small values of $\Delta_1$ for the XXZ model and that, in the latter regimes, $J_2\sim 2.0$ gives the values that are closest to $\pi/2$. 

In Appendix~\ref{app::param_scan}, we show as density plots the normalized differences $|\Gamma_\psi - \pi/2|/(\pi/2)$ as functions of $J_2$, $h^z$ for the XYZ model and as functions of $J_2$, $\Delta_1$ for the XXZ model. We compare them with results for the normalized differences $|\bar r -\bar r_{\rm GOE}|/\bar r_{\rm GOE}$ computed in the same parameter regimes. Those plots provide a more complete picture of the regimes in which the results for the model Hamiltonians are closest to the RMT predictions and about the sensitivity of the results obtained for $\bar r $ vs that of the results obtained for $\Gamma_\psi$. 

As a side remark, we note that for the system sizes considered in Fig.~\ref{fig:coefficients}, $\Gamma_\psi$ is slightly greater than $\pi/2$ even in the maximally chaotic regime of both models. In Appendix~\ref{sec:app_scaling}, we carry out finite-size scaling analyses of the normalized difference $|\Gamma_\psi-\pi/2|/(\pi/2)$ for the XYZ model in chains with both periodic and open boundary conditions. We show that the normalized differences decrease with increasing system size, but they decrease more slowly than for eigenstates of random matrices and than for pure random states with coefficients that are drawn from a normal distribution.


\subsection{Average eigenstate entanglement entropy} \label{sec:chaos:entropy}

To close this section on the maximally chaotic regime, we explore the behavior of the eigenstate entanglement entropy $S_A$ of Hamiltonian eigenstates, see Eq.~\eqref{Neumann.entropy}. We focus on bipartitions in two halves, i.e., we set the subsystem fraction to $f=1/2$, and study the same parameter regimes of the Hamiltonians as in Secs.~\ref{sec:levelspacing} and~\ref{sec:coeff}.

\begin{figure}[!b]
\includegraphics[width=0.98\columnwidth]{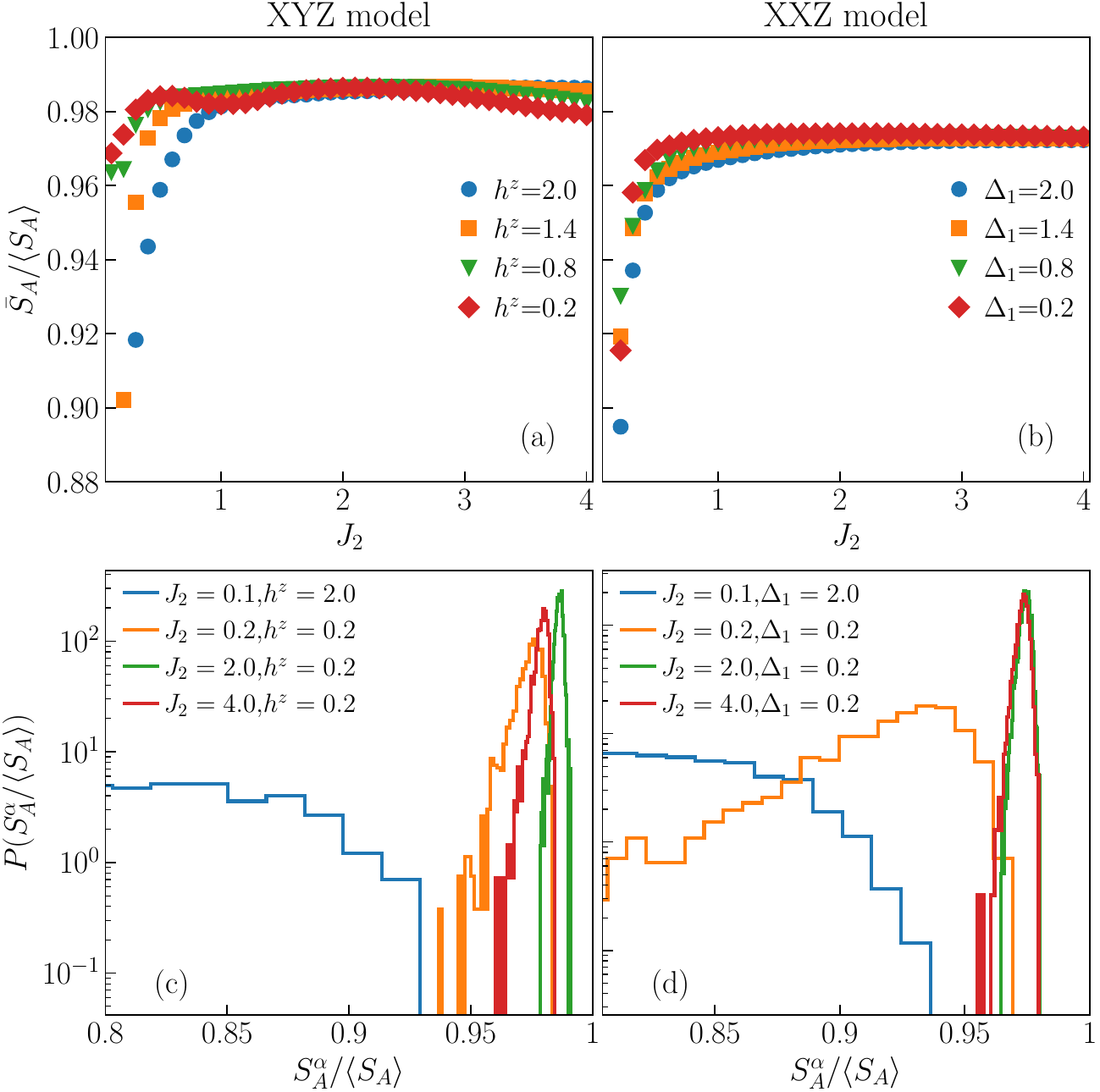}
\caption{Scaled average eigenstate entanglement entropy $\bar S_A/\langle S_A\rangle$, obtained after tracing out 1/2 of the chain, for the same models and parameters as in Fig.~\ref{fig:levelspac}. [(a) and (b)] Results for the XYZ ($L=18$) and XXZ ($L=20$) models, respectively. The averages are calculated taking 100 midspectrum eigenstates from each symmetry block, and the Page value $\langle S_A \rangle$ is taken to be the two leading terms in Eq.~\eqref{eq:page_inf}. [(c) and (d)] Distributions $P(S_A^\alpha/\langle S_A\rangle)$, of the eigenstate entanglement entropies $S_A^\alpha$ used when computing the averages, for four sets of Hamiltonian parameters.}
\label{fig:aveentall}
\end{figure}

Figures~\ref{fig:aveentall}(a) and~\ref{fig:aveentall}(b) show the average entanglement entropy $\bar S_A$ in both models for the same Hamiltonian parameters as those in the study of the average level spacing $\bar r$ in Figs.~\ref{fig:levelspac}(a) and~\ref{fig:levelspac}(b). We average $S_A$ over 100 midspectrum eigenstates from each symmetry block and divide the results by the two leading terms in Page's prediction, see $\langle S_A \rangle$ in Eq.~\eqref{eq:page_inf}. A comparison between Fig.~\ref{fig:aveentall}(a) [Fig.~\ref{fig:aveentall}(b)] and Fig.~\ref{fig:levelspac}(a) [Fig.~\ref{fig:levelspac}(b)] reveals a similar trend between the deviation of $\bar S_A$ from Page's leading-order prediction and the deviation of $\bar r$ from the RMT prediction. We do note that the deviations in the average entanglement entropy are more pronounced at small $J_2$ in both models. This is more prominent for the XXZ model in Fig.~\ref{fig:aveentall}(b), for which we find that $\bar S_A$ at moderate $\Delta_1$ increases with $J_2$ even in the regime $J_2>1$. This observation motivate us in the next section to set $J_2=2.0$ to carry out the scaling analyses.

In Appendix~\ref{app::param_scan}, we show density plots of the normalized differences between $\bar S_A$ and Page's leading-order prediction for $\langle S_A \rangle$, $|\bar S_A -\langle S_A \rangle|/\langle S_A \rangle$, as functions of $J_2$, $h^z$ for the XYZ model and as functions of $J_2$, $\Delta_1$ for the XXZ model. Like the results for $\bar r$ and $\Gamma_\psi$, those plots provide a more complete picture of where the agreement between the averages over Hamiltonian eigenstates and the theoretical expectations are closest.

In Figs.~\ref{fig:aveentall}(c) and \ref{fig:aveentall}(d), we show exemplary distributions $P(S_A^\alpha/\langle S_A\rangle)$ of the eigenstate entanglement entropies $S_A^\alpha$ used when computing the averages for four sets of Hamiltonian parameters for the XYZ and XXZ models, respectively. One can see that the closer the average $\bar S_A$ is to Page's result the narrower is the distribution (or, what is the same, higher averages come from narrower distributions). Narrow distributions of eigenstate expectation values of few-body observables are a hallmark of eigenstate thermalization and, hence, of quantum chaos~\cite{dalessio_kafri_16}. It is remarkable that the same applies to the entanglement entropy of 1/2 of the system, which is a multibody observable. The most chaotic regime in Fig.~\ref{fig:aveentall}(c) [Fig.~\ref{fig:aveentall}(d)] is found to be the regime of $J_2\sim 2$ and small $h^z$ ($\Delta_1$) in the XYZ (XXZ) model.

Summarizing our analysis in this section, we showed that all three quantities under investigation, the level spacing ratio $\bar r$, the statistics of the eigenstate coefficients $c_m^\alpha$, and the eigenstate entanglement entropy $S_A$ yield consistent information about the degree of quantum chaos in different regimes of model parameters, with the latter two depending more strongly on the values of the model parameters.

\section{Scaling of the Average Entanglement Entropy} \label{sec:entro}

We now turn our attention to the main subject of our work, i.e., the behavior of the average entanglement entropy of midspectrum eigenstates. We are interested in exploring their scaling with the system size and with the fraction of midspectrum eigenstates taken to carry out the average. Motivated by the results from Sec.~\ref{sec:chaos} and Appendix~\ref{app::param_scan}, we select model parameters for which the system is found to be maximally chaotic. Specifically, we set $J_2=2.0$ and $h^z = 0.2$ for the XYZ model and $J_2=2.0$ and $\Delta_1 = 0.2$ for the XXZ model.

\begin{figure}[!t]
\includegraphics[width=0.98\columnwidth]{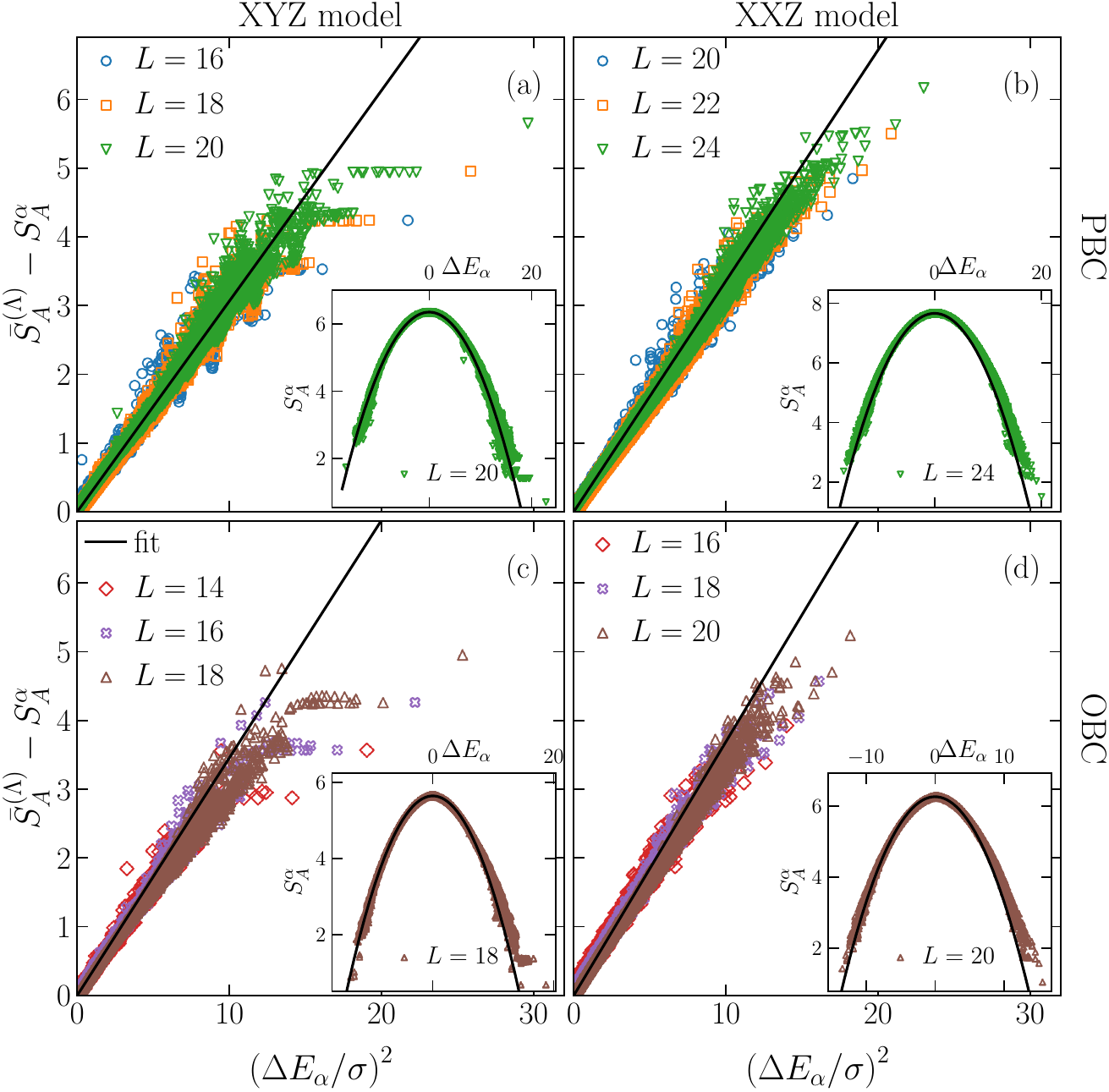}
\caption{Eigenstate entanglement entropies $S^\alpha_A$ at $f=1/2$, in all symmetry blocks, as functions of $\Delta E_\alpha$ (see text). The main panels show the deviation from the average $\bar S_A^{(\Lambda)}$ vs $(\Delta E_\alpha/\sigma)^2$ [$\sigma^2$ is defined in Eq.~\eqref{eq:sigma}] for different system sizes, while the insets show $S^\alpha_A$ vs $\Delta E_\alpha$ for the largest system size computed in each case. We calculate $\bar S_A^{(\Lambda)}$ using $\Lambda=100$ midspectrum eigenstates from each quasimomentum sector. [(a) and (b)] Results for chains with periodic boundary conditions (PBCs) for the XYZ and XXZ models, respectively. [(c) and (d)] Results for chains with open boundary conditions (OBCs) for the XYZ and the XXZ models, respectively. The solid lines in the insets show fits of the eigenstate entanglement entropies to Eqs.~\eqref{eq:concaveS} and~\eqref{eq:concaveC}, in which the only fitting parameter is $c$. The values of $\bar S_A^{(\Lambda)}$ and $c^2$ for the curves shown are: (a) $\bar S_A^{(\Lambda)}=6.35$ and $c^2=3.26$, (b) $\bar S_A^{(\Lambda)}=7.66$ and $c^2=2.98$, (c) $\bar S_A^{(\Lambda)}=5.65$ and $c^2=2.89$, and (d) $\bar S_A^{(\Lambda)}=6.27$ and $c^2=2.71$. (For the fits, we use the entanglement entropies of the central 50\% of the energy eigenstates.) The solid lines in the main panels show $(\Delta E_\alpha)^2/(c^2\,\sigma^2)$, where $c$ comes from the fits in the insets, and $\sigma$ is computed for each system size.}
\label{fig:entropyvsE}
\end{figure}

It is well known that the eigenstate entanglement entropy of local Hamiltonians is a concave function of the eigenenergies, with a maximum in the middle of the energy spectrum, namely, about $\bar E={\text{Tr}}(\hat H)/\mathcal{D}$, with $\mathcal{D}$ being the dimension of the entire Hilbert space~\cite{deutsch_li_2013, beugeling_andreanov_15, garrison_grover_18, murthy_srednicki_19a, miao_barthel_21, Haque_Khaymovich_2022}. This property is illustrated in the insets in Fig.~\ref{fig:entropyvsE} for our two models, using periodic (top panels) and open (bottom panels) boundary conditions. In those insets, we plot $S_A^\alpha$ vs $\Delta E_\alpha=E_\alpha-\bar{\mathrm E}$ for all symmetry blocks, where $\bar {\mathrm E}$ is the average energy in the symmetry block to which $E_\alpha$ belongs ($\bar {\mathrm E}\rightarrow \bar E$ with increasing system size, as shown in Fig.~\ref{fig:app:escaling}). Consequently, in finite systems, the average eigenstate entanglement entropy depends on the number of states used to carry out the average. In large systems, as we show in what follows, this translates into a dependence of the subleading $O(1)$ term on the fraction of eigenstates used to carry out the averages. 

About the maximum value of $S^\alpha_A$, which can be computed as an average $\bar S_A^{(\Lambda)}$ over a fixed number $\Lambda$ of midspectrum eigenstates, the insets in Fig.~\ref{fig:entropyvsE} (containing all symmetry blocks) make apparent that one can write the eigenstate entanglement entropy $S^\alpha_A$ as
\begin{equation}\label{eq:concaveS}
S^\alpha_A=\bar S_A^{(\Lambda)} - C^2 (\Delta E_\alpha)^2,
\end{equation}
where $C$, in general, depends on the Hamiltonian parameters and the system size. We find that 
\begin{equation}\label{eq:concaveC}
C\simeq \frac{1}{c\,\sigma},
\end{equation}
where $\sigma^2$ is the variance of the energy of the {\it entire} energy spectrum 
\begin{equation}\label{eq:sigma}
\sigma^2 = \frac{1}{\mathcal{D}}\sum_\alpha (E_\alpha-\bar E)^2,
\end{equation}
and $c$ is independent of the system size. This is demonstrated by the data collapse seen in the main panels in Fig.~\ref{fig:entropyvsE}, in which we plot $\bar S_A^{(\Lambda)} -S^\alpha_A$ vs $(\Delta E_\alpha/\sigma)^2$ for the entire energy spectrum in chains with different sizes. Remarkably, the fits of the numerical results in insets in Fig.~\ref{fig:entropyvsE} to Eqs.~\eqref{eq:concaveS} and~\eqref{eq:concaveC} show that those equations provide an excellent description of $S^\alpha_A$ over most of the energy spectrum. This is further confirmed by the data collapse in the main panels, which occurs about the fits.

In this work we carry out two types of averages for the eigenstate entanglement entropy. The first one was already mentioned before, namely we average over a fixed number $\Lambda$ of midspectrum eigenstates. This fixed number, as one increases the system size, corresponds to an exponentially vanishing fraction relative to the total number of eigenstates. The energy eigenstates in that average have eigenenergies $E_\alpha \rightarrow \bar {\mathrm E}$ exponentially fast with increasing system size. In the second type of average, we use a finite fraction $\nu$ of midspectrum eigenstates, $\nu\in (0, 1]$, and we denote the average as $\bar S_A^{(\nu)}$. It is important to emphasize that, when computing the averages, we define the ``midspectrum eigenstates'' for each symmetry block separately, i.e., they are the eigenstates whose eigenenergies are closest to the mean energy $\bar {\mathrm E}$ in each symmetry block.

Using Eqs.~\eqref{eq:concaveS} and~\eqref{eq:concaveC}, we can estimate the dependence of the average eigenstate entanglement entropy on the fraction $\nu$ of midspectrum eigenstates used to compute the averages. Let us begin by noticing that, for models with few-body interactions (our interest here), the density of states is Gaussian about the mean energy $\bar E$. Introducing a new variable ${\mathcal E}=E-\bar E$, we can write
\begin{equation} \label{eq:dos}
    \rho ({\mathcal E}) = \mathcal{D} \frac{1}{\sqrt{2 \pi } \sigma } e^{-\frac{{\mathcal E}^2}{2\sigma ^2}},\quad \text{so that}\quad \int _{-\infty}^{\infty} \rho({\mathcal E}) d {\mathcal E} = \mathcal{D},
\end{equation}
and has a variance $\sigma^2\propto L$~\cite{brody_81, mondaini_rigol_17}. 

Using Eq.~\eqref{eq:dos}, we can write
\begin{equation}\label{eq:nu_deriv}
    \nu = \frac{1}{\mathcal{D}} \int _{-{\mathcal E}_\nu} ^ {{\mathcal E}_\nu} \rho({\mathcal E}) d{\mathcal E}= \erf \left(\frac{{\mathcal E}_\nu}{\sqrt{2}\sigma}\right),
\end{equation}
which shows that ${\mathcal E}_\nu \propto \sigma \propto \sqrt{L}$ for $\nu \in (0, 1)$. The average entanglement entropy over this energy window is then given by
\begin{equation}
\bar{\mathcal{S}} ^{(\nu)}_A = \frac{1}{\nu \mathcal{D}} \int _{-{\mathcal E}_\nu}^{{\mathcal E}_\nu} \rho({\mathcal E})\, S_A({\mathcal E})\, d{\mathcal E}.
\end{equation}
Inserting Eqs.~\eqref{eq:concaveS},~\eqref{eq:concaveC}, and \eqref{eq:dos} into the equation above, one obtains
\begin{eqnarray} \label{eq:ent_av_deriv}
    \bar{\mathcal{S}}^{(\nu)}_A &=& \frac{1}{\nu \mathcal{D}} \int ^{{\mathcal E}_\nu}_{-{\mathcal E}_\nu} \frac{\mathcal{D}}{\sqrt{2\pi} \sigma} e^{-\frac{{\mathcal E}^2}{2 \sigma ^2}} \left[ \bar S_A^{(\Lambda)} - \frac{{\mathcal E}^2}{c^2\sigma^2} \right] d{\mathcal E} = \nonumber\\
    && = \bar S_A^{(\Lambda)} - \frac{1}{c^2} \left[ 1 - \sqrt{\frac{2}{\pi}}\, \frac{{\mathcal E}_\nu }{\nu\,\sigma} \, e^{-\frac{{\mathcal E}^2_\nu}{2\sigma^2}}\right].
\end{eqnarray}
From Eq.~\eqref{eq:nu_deriv}, we know that ${\mathcal E}_\nu / \sigma = \sqrt{2} \erf ^{-1} (\nu)$. Hence, we find that the concave functional form of $S_A^\alpha$ vs $\Delta E_\alpha$ with a maximum in the middle of the spectrum results in an $O(1)$ correction of the average over a finite fraction $\nu$ of the energy eigenstates when compared to the maximal result
\begin{equation}\label{eq:Snuphenomen}
     \bar{\mathcal{S}}^{(\nu)}_A = \bar S_A^{(\Lambda)} - \frac{1}{c^2} \left[1-\frac{2}{\sqrt{\pi}} \frac{\erf ^{-1} (\nu)}{\nu} e^{-\left[\erf ^{-1} (\nu) \right]^2}\right].
\end{equation}

In Secs.~\ref{sec:xyz} and~\ref{sec:xxz}, we report numerical results for the average entanglement entropy of a fraction $\nu$ of midspectrum eigenstates in nonintegrable XYZ and XXZ chains, respectively.  We show that, after computing $\bar S_A^{(\Lambda)}$ and fitting $c^2$, Eq.~\eqref{eq:Snuphenomen} describes the results for $\bar{\mathcal{S}}^{(\nu)}_A$ in our numerical calculations. When meaningful, in Secs.~\ref{sec:xyz} and~\ref{sec:xxz}, we will also report results for the maximal and minimal eigenstate entanglement entropies within the set of states over which the average is carried out. In contrast to the average eigenstate entanglement entropies, the maximal and minimal eigenstate entanglement entropies are not averaged over symmetry blocks. We select the maximal and minimal over the entire set containing all sectors and refer to them as the ``outlier'' eigenstate entanglement entropies. They bound the results for the entanglement entropies in the set considered.

All the results in the main text correspond to the subsystem fraction $f=1/2$.
Results for other system fractions are shown in Appendix~\ref{sec:fneq05}.

\subsection{XYZ model}\label{sec:xyz}

\begin{figure}[!t]
\includegraphics[width=0.834\columnwidth]{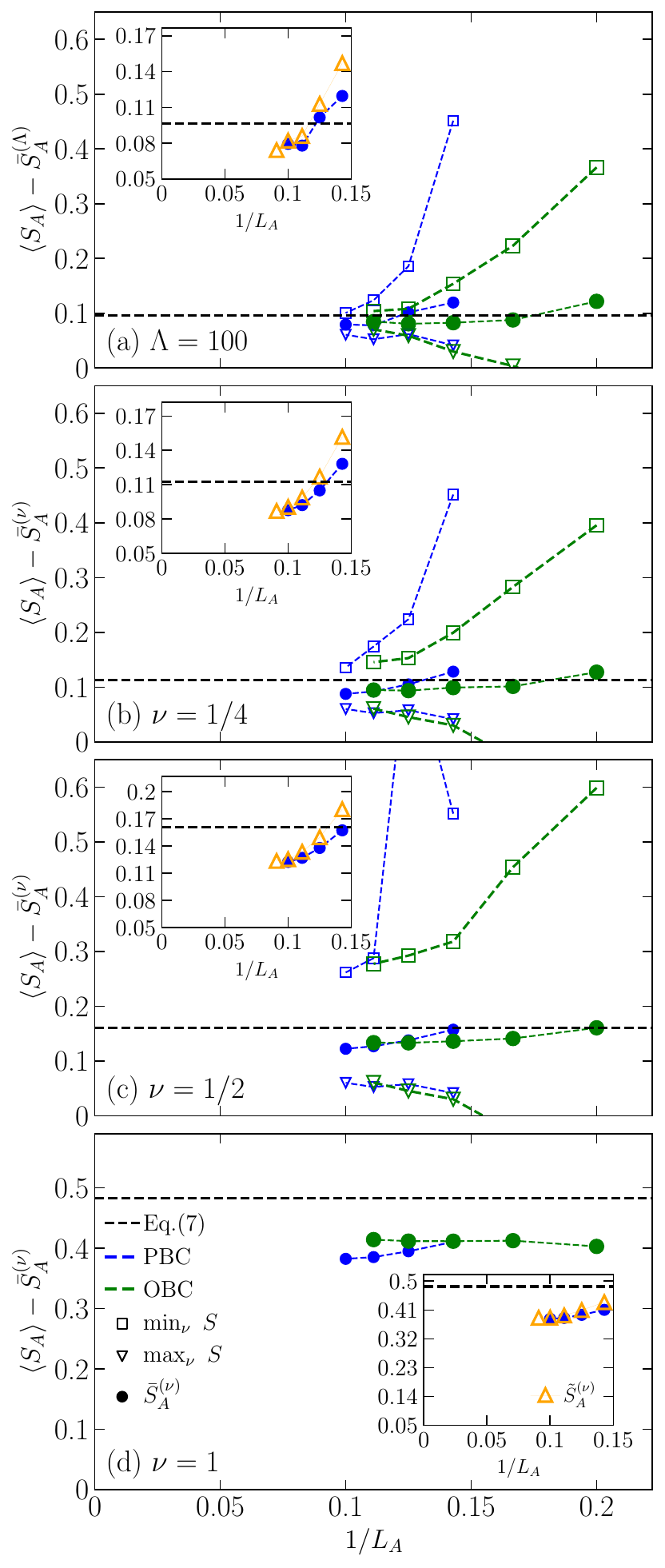}
\vspace{-0.3cm}
\caption{Finite-size scaling analysis for the XYZ model. The deviations of the average eigenstate entanglement entropy $\bar S_A^{(\nu)}$ (filled symbols), and the outlier eigenstate entanglement entropies (open symbols), from the exact result for the average over random pure states $\langle S_A \rangle$ [Eq.~\eqref{eq:page_finite}] are plotted vs the inverse subsystem size $L_A=L/2$. Results are shown for chains with both periodic and open boundary conditions. The insets show the average eigenstate entanglement entropy in systems with periodic boundary conditions including all quasimomentum sectors, $\bar S_A^{(\nu)}$ (same as in the main panels), and only including the $k=0$ and $\pi$ sectors, $\Tilde{S}^{(\nu)} _A$ (``real'' sectors for which we can diagonalize the largest chains). The averages are carried out over (a) $\Lambda=100$ ($\nu=0^+$), (b) $\nu=1/4$, (c) $\nu=1/2$, and (d) the entire spectrum ($\nu=1$). The horizontal dashed lines show the predictions of Eq.~\eqref{eq:nu}.}
\label{fig:entropyscalXYZ}
\end{figure}

In order to carry out a finite-size scaling analysis for the XYZ model, in this section we compute the average entanglement entropy of energy eigenstates in chains with different sizes $L$. We then subtract the numerical results obtained from the prediction by Page~\cite{page_1993b} for the average over random pure states in a full Hilbert space $\mathcal{H} = \mathcal{H_A} \otimes \mathcal{H}_B$, with corresponding dimensions ${\cal D}_A = 2^{L_A}$ and ${\cal D}_B = 2^{L-L_A}$, which has the form~\cite{bianchi_hackl_22}:
\begin{equation}\label{eq:page_finite}
{\small \langle S _A \rangle = 
    \left\{ \begin{array}{ll}
         \Psi(\mathcal{D}_A\mathcal{D}_B + 1)- \Psi(\mathcal{D}_B+1) - \frac{\mathcal{D}_A - 1}{2\mathcal{D}_B},& \mathcal{D}_A \leq \mathcal{D}_B \\
         \Psi(\mathcal{D}_A\mathcal{D}_B + 1)- \Psi(\mathcal{D}_A+1) - \frac{\mathcal{D}_B - 1}{2\mathcal{D}_A},& \mathcal{D}_A > \mathcal{D}_B
    \end{array} \right.,}
\end{equation}
where $\Psi (z) = (\ln{\Gamma (z)})' = \Gamma ' (z) / \Gamma (z)$ is the digamma function. At subsystem fraction $f = L_A / L$, the two leading terms for Eq.~\eqref{eq:page_finite} are given in Eq.~\eqref{eq:page_inf}.

In Fig.~\ref{fig:entropyscalXYZ}, we show the finite-size scaling of the differences $\langle S_A \rangle - \bar S_A^{(\nu)}$ (filled symbols) for the average over: (a) $\Lambda=100$ ($\nu=0^+$), (b) $\nu=1/4$, (c) $\nu=1/2$, and (d) all eigenstates ($\nu=1$). The open symbols in Fig.~\ref{fig:entropyscalXYZ} show the differences for the outlier eigenstate entanglement entropies. In the main panels we show results for periodic and open boundary conditions, while in the insets we show results for periodic boundary conditions for averages over all quasimomentum sectors (as shown in the main panels) and only over $k=0$ and $\pi$ (for which the largest system sizes can be diagonalized). We also show, as horizontal dashed lines, the prediction of Eq.~\eqref{eq:nu} for the specific value of $\nu$ under consideration.

In Fig.~\ref{fig:entropyscalXYZ}(a) and its inset, for $\Lambda=100$ midspectrum eigenstates, one can see that with increasing system size the difference between the average over random pure states and the average over Hamiltonian eigenstates appears to saturate at a small $O(1)$ number that is smaller than, but likely of the order of, $0.1$. This $O(1)$ number seems to be slightly smaller than the one predicted by Eq.~\eqref{eq:nu}, which, as mentioned in the introduction, is identical to the result for the average over random pure states with fixed magnetization $s^z=0$ at $f=1/2$ [$n=1/2$ and $f=1/2$ in Eq.~\eqref{eq:nleading}].

\begin{figure}[!t]
\includegraphics[width=0.9\columnwidth]{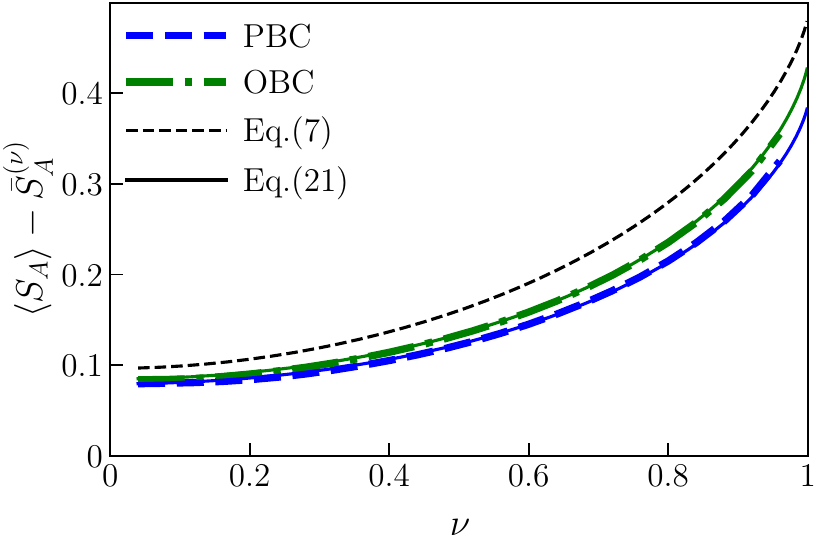}
\vspace{-0.1cm}
\caption{Average eigenstate entanglement entropy vs $\nu$ for the XYZ model. The differences between the average eigenstate entanglement entropy $\bar S_A^{(\nu)}$ and the average over random pure states (thick lines) are plotted as functions of the fraction $\nu$ of midspectrum Hamiltonian eigenstates included in the average. We show results for a chain with $L=20$ with PBCs and for a chain with $L=18$ with OBCs. The thin solid lines overlapping with the numerical results are the predictions of Eq.~\eqref{eq:Snuphenomen}, with the values of $\bar S_A^{(\Lambda)}$ and $c^2$ taken from the curves shown in the insets in Figs.~\ref{fig:entropyvsE}(a) and~\ref{fig:entropyvsE}(c). The thin dashed line is the function $\bar{\text{S}}_A^{(\nu)}$ from Eq.~\eqref{eq:nu}.}
\label{fig:entropyvsnuXYZ}
\end{figure}

In Figs.~\ref{fig:entropyscalXYZ}(b)--\ref{fig:entropyscalXYZ}(d), one can see that with increasing system size $\langle S_A \rangle - \bar S_A^{(\nu)}$ approaches a nonzero $O(1)$ constant whose value increases with $\nu$. In Fig.~\ref{fig:entropyvsnuXYZ}, we plot the difference $\langle S_A \rangle - \bar S_A^{(\nu)}$ vs $\nu$ for a chain with periodic boundary conditions (PBCs) ($L=20$) and for a chain with open boundary conditions (OBCs) ($L=18$). In both cases we find the difference to be in excellent agreement with the prediction of Eq.~\eqref{eq:Snuphenomen}, with the values of $\bar S_A^{(\Lambda)}$ and $c^2$ taken from the curves shown in the insets in Figs.~\ref{fig:entropyvsE}(a) and~\ref{fig:entropyvsE}(c). Since the results for large systems with PBCs and OBCs are expected to agree up to corrections that vanish in the thermodynamic limit, the fact that the PBC and OBC results are slightly different makes clear that they still suffer from finite-size effects and are expected to decrease (likely only slightly) if larger system sizes are considered.

Like in Fig.~\ref{fig:entropyscalXYZ}(a), in Figs.~\ref{fig:entropyscalXYZ}(b)--\ref{fig:entropyscalXYZ}(d) as well as in Fig.~\ref{fig:entropyvsnuXYZ}, we consistently find our numerical results to be below the predictions from Eq.~\eqref{eq:nu}. The results for the outliers in Figs.~\ref{fig:entropyscalXYZ}(a)--\ref{fig:entropyscalXYZ}(c) further show the range of values over which the average is carried out. Figure~\ref{fig:entropyscalXYZ}(a) shows that, strikingly, the least entangled Hamiltonian eigenstates among the $\Lambda=100$ midspectrum eigenstates, the ones with the largest $\langle S_A \rangle - S^\alpha_A$, are already at the value for the average predicted by Eq.~\eqref{eq:nu}. This strengthens our expectation that $\bar S_A^{(\nu)}$ will be greater than the prediction from Eq.~\eqref{eq:nu} for $\nu=0^+$ [for an $O(1)$ number of eigenstates] in the thermodynamic limit. Furthermore, the results in Fig.~\ref{fig:entropyvsnuXYZ} show that $\bar{\text{S}}_A^{(\nu)}$ from Eq.~\eqref{eq:nu} does not capture the functional dependence of $\langle S_A \rangle - \bar S_A^{(\nu)}$ vs $\nu$ observed in the numerical results. Given our Eq.~\eqref{eq:Snuphenomen}, we expect that $\bar S_A^{(\nu)}$ will in general depend on the model Hamiltonian under consideration through a potentially nonuniversal $O(1)$ term in $\bar S_A^{(\Lambda)}$ and the constant $c$. 

In Appendix~\ref{sec:app_scaling}, we report scalings similar to the one in Fig.~\ref{fig:entropyscalXYZ}(a) obtained for other Hamiltonian parameters across and beyond the maximally chaotic regime. They show that the results in Fig.~\ref{fig:entropyscalXYZ}(a) are robust against changes in the Hamiltonian parameters and, hence, that our findings and conclusions are not a consequence of a fine tuning of parameters for the specific model under consideration.

\subsection{XXZ model}\label{sec:xxz}

\begin{figure}[!ht]
\includegraphics[width=0.852\columnwidth]{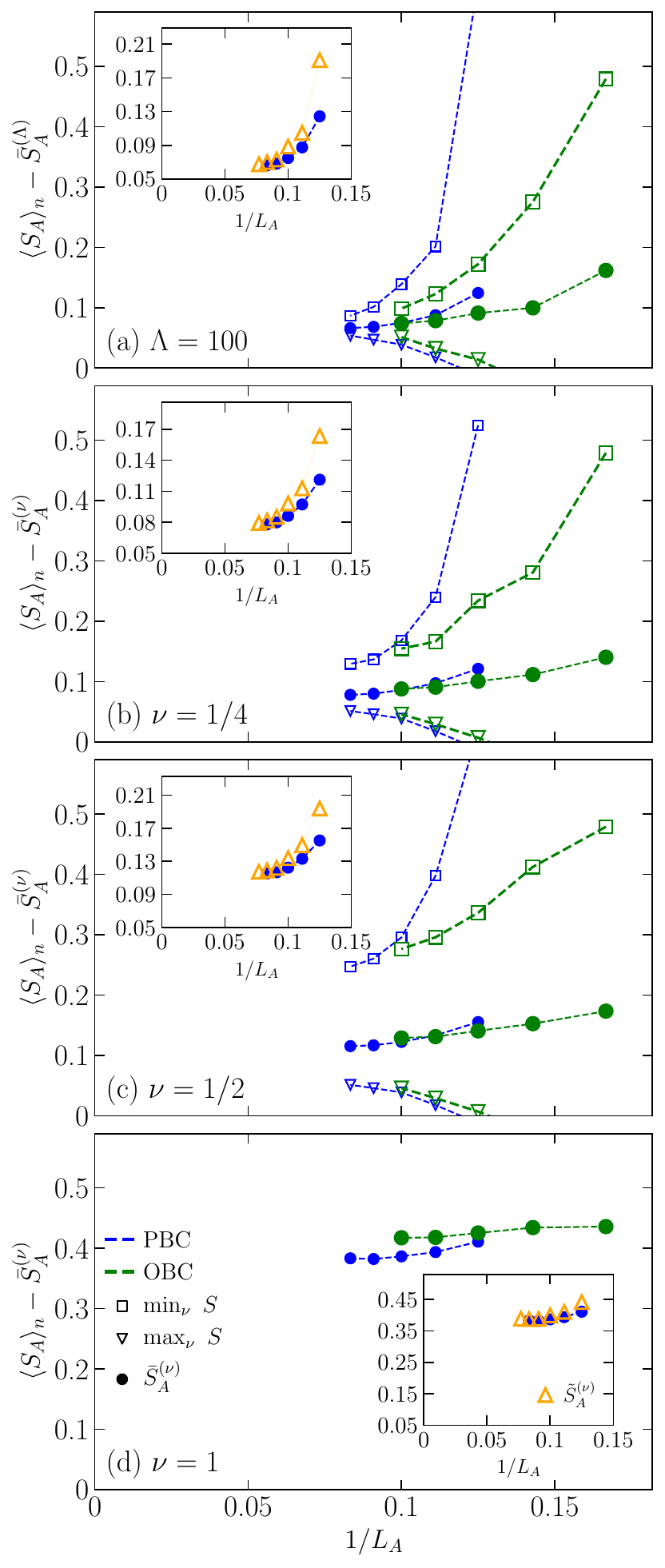}
\vspace{-0.3cm}
\caption{Finite-size scaling analysis for the XXZ model. The deviations of the average eigenstate entanglement entropy $\bar S_A^{(\nu)}$ (filled symbols), and the outlier eigenstate entanglement entropies (open symbols), from the exact result for random pure states $\langle S_A\rangle_n$ [Eq.~\eqref{eq:page_n}] are plotted vs the inverse subsystem size $L_A=L/2$. Results are shown for chains with both periodic and open boundary conditions. The insets show the average eigenstate entanglement entropy in systems with periodic boundary conditions including all quasimomentum sectors, $\bar S_A^{(\nu)}$ (same as in the main panels), and only including the $k=0$ and $\pi$ sectors, $\Tilde{S}^{(\nu)} _A$ (``real'' sectors for which we can diagonalize the largest lattices). The averages are carried out over (a) $\Lambda=100$ ($\nu=0^+$), (b) $\nu=1/4$, (c) $\nu=1/2$, and (d) the entire spectrum ($\nu=1$).}
\label{fig:entropyscalXXZ}
\end{figure}

In order to carry out the finite-size scaling analysis for the XXZ model, we subtract our numerical results from the exact result obtained when averaging over random states in finite-dimensional Hilbert spaces at fixed zero magnetization, or, in the spinless fermions language, at fixed half filling. Using the latter (more convenient) language, the Hilbert space $\mathcal{H}^{(N)}$ of a system with $N$ spinless fermions in $L$ sites is a direct sum of tensor products of Hilbert spaces in subsystems $A$ (with $L_A$ sites and $N_A$ fermions) and $B$ (with $L_B=L-L_A$ sites and $N_B=N-N_A$ fermions),
\begin{equation}
    \mathcal{H}^{(N)} = \bigoplus _{N_A=0}^{\min (N, L_A)} \left(\mathcal{H}_A^{(N_A)} \otimes \mathcal{H}_B^{(N-N_A)}\right) \;,
\end{equation}
where the latter equation assumes $N \leq L/2$, and we define $n=N/L$. The dimension of the total Hilbert space $\mathcal{H}^{(N)}$ is ${\cal D}_N = {L \choose N}$, and of the Hilbert spaces $\mathcal{H}_i^{(N_i)}$, with $i=A,B$, are
\begin{equation}
    \mathcal{D} _ i (N_i) = \dim \mathcal{H}_i^{(N_i)} = {L_i\choose N_i}.
\end{equation}
The average entanglement entropy of random pure states in a system with $L$ sites and $N$ spinless fermions, after tracing out $L-L_B$ sites, takes the form~\cite{bianchi_dona_19, bianchi_hackl_22}
\begin{eqnarray} \label{eq:page_n}
        &&  \langle S_A\rangle _n = \sum _{N_A=0} ^ {\min (N, L_A)} \frac{\mathcal{D} _A (N_A) \mathcal{D} _B (N_B)}{\mathcal{D}_N} \times \\
        &&\qquad\  \left[ \langle S_A\rangle + \Psi (\mathcal{D} _N + 1) -\Psi (\mathcal{D}_A (N_A) \mathcal{D}_B (N_B) + 1)  \right],\nonumber
\end{eqnarray}
in which $\langle S_A \rangle$ is given by Eq.~\eqref{eq:page_finite}. At subsystem fraction $f = L_A / L$, the leading terms for Eq.~\eqref{eq:page_n} are given in Eq.~\eqref{eq:nleading}. Our focus in this section is $f=1/2$ at half filling $n=1/2$ (corresponding to zero magnetization).

\begin{figure}[!t]
\includegraphics[width=0.9\columnwidth]{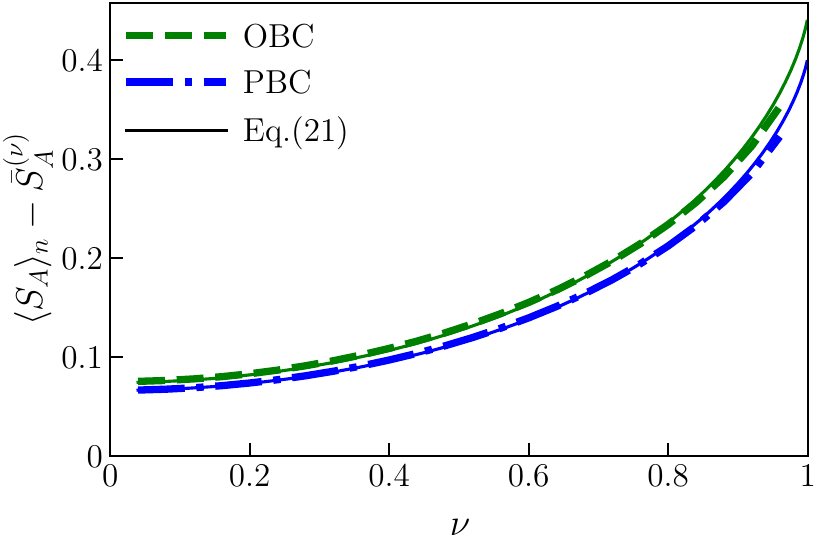}
\vspace{-0.1cm}
\caption{Average eigenstate entanglement entropy vs $\nu$ for the XXZ model. The differences between the average eigenstate entanglement entropy $\bar S_A^{(\nu)}$ and the average over random pure states (thick lines) are plotted as functions of the fraction $\nu$ of midspectrum Hamiltonian eigenstates included in the average. We show results for a chain with $L=24$ with PBCs and for a chain with $L=20$ with OBCs. The thin solid lines overlapping with the numerical results are the predictions of Eq.~\eqref{eq:Snuphenomen}, with the values of $\bar S_A^{(\Lambda)}$ and $c^2$ taken from the curves shown in the insets in Figs.~\ref{fig:entropyvsE}(b) and~\ref{fig:entropyvsE}(d).}
\label{fig:entropyvsnuXXZ}
\end{figure}

The finite-size scalings of the differences $\langle S_A \rangle_n - \bar S_A^{(\nu)}$ are shown in Fig.~\ref{fig:entropyscalXXZ}, for the same number $\Lambda=100$ of midspectrum eigenstates [Fig.~\ref{fig:entropyscalXXZ}(a)] and fractions of midspectrum eigenstates [Figs.~\ref{fig:entropyscalXXZ}(b)--\ref{fig:entropyscalXXZ}(d)] as those in the respective panels in Fig.~\ref{fig:entropyscalXYZ}. The similarity between the finite-size scaling results in Fig.~\ref{fig:entropyscalXXZ} and in Fig.~\ref{fig:entropyscalXYZ} is striking. They suggest that for the models considered the deviations of $\bar S_A^{(\nu)}$, for any given value of $\nu$, from the result for the average over random states is nearly independent of whether the Hamiltonian exhibits or not $U(1)$ symmetry (particle-number conservation). We emphasize that this is the case despite the fact that the $O(1)$ subleading term in $\langle S_A \rangle$ (from which the XYZ results are subtracted) and in $\langle S_A \rangle_n$ (from which the XXZ results are subtracted) are different, see Eqs.~(\ref{eq:nleading}) and~(\ref{eq:page_inf}).

In Fig.~\ref{fig:entropyvsnuXXZ}, we plot $\langle S_A \rangle_n - \bar S_A^{(\nu)}$ vs $\nu$ for a chain with PBCs ($L=24$) and for a chain with OBCs ($L=20$). The plots are qualitatively and quantitatively similar to the ones seen for the XYZ model in Fig.~\ref{fig:entropyvsnuXYZ}. Also like in Fig.~\ref{fig:entropyvsnuXYZ}, we find that $\langle S_A \rangle_n - \bar S_A^{(\nu)}$ vs $\nu$ is in excellent agreement with the prediction of Eq.~\eqref{eq:Snuphenomen}, with the values of $\bar S_A^{(\Lambda)}$ and $c^2$ taken from the curves shown in the insets in Figs.~\ref{fig:entropyvsE}(b) and~\ref{fig:entropyvsE}(d). We note that, both in Figs.~\ref{fig:entropyvsnuXYZ} and~\ref{fig:entropyvsnuXXZ}, the average over up to $\sim 20\% $ of the midspectrum eigenstates barely changes the result from that at $\nu=0^+$. Those fractions can be used in numerical calculations to reduce fluctuations in the average over the entanglement entropy of Hamiltonian eigenstates associated to finite-size effects, while still producing results that are close to those at $\nu=0^+$.

In Appendix~\ref{sec:app_scaling}, we report scalings similar to the one in Fig.~\ref{fig:entropyscalXXZ}(a) obtained for other Hamiltonian parameters across and beyond the maximally chaotic regime. They show that the results in Fig.~\ref{fig:entropyscalXXZ}(a) are robust against changes in the Hamiltonian parameters and, hence, that our findings and conclusions are not a consequence of a fine tuning of the parameters for the specific model under consideration.

\section{Summary and discussion} \label{sec:conclusions}

We carried out a state-of-the-art computational study of the $O(1)$ subleading corrections to the leading volume-law term of the average entanglement entropy of midspectrum eigenstates in nonintegrable spin-1/2 XYZ and XXZ chains. We focused on the subsystem fraction $f=1/2$, and for the XXZ chain [which has $U(1)$ symmetry] we focused on the zero magnetization sector. 

For a fixed number $\Lambda$ of midspectrum Hamiltonian eigenstates in the average ($\Lambda=100$, i.e., the fraction of Hamiltonian eigenstates $\nu=0^+$), we found indications that the average eigenstate entanglement entropy differs by a small $O(1)$ number from the prediction for the average over random states in the thermodynamic limit. The magnitude of the difference was $\lesssim 0.1$ both for the XYZ [does not have $U(1)$ symmetry] and the XXZ [has $U(1)$ symmetry] chains. While the magnitude of the difference was found to be similar in the absence or presence of the $U(1)$ symmetry, it is important to emphasize that the average over random states exhibits an $O(1)$ term that {\it does depend} on whether the average is carried out over states in which the magnetization is fixed or not~\cite{bianchi_hackl_22}. 

We also found indications that the fixed-$\Lambda$ average entanglement entropy of eigenstates of the XYZ model differs from Huang's prediction (the former is greater for the models and parameters considered here) in the thermodynamic limit. Huang's $O(1)$ correction at $\nu=0^+$ is identical to the $O(1)$ ``mean-field'' correction derived in Ref.~\cite{vidmar_rigol_17} for the average over random pure states at fixed particle number (fixed magnetization in the spin language). In Appendix~\ref{sec:EfromN}, we show that a finite-size scaling analysis of the difference between the average entanglement entropy of eigenstates of the XYZ model and the analytic prediction for the average over random pure states at fixed zero magnetization also indicates that those two averages exhibit an $O(1)$ difference. This is understandable as the energy conservation constraint in quantum-chaotic local Hamiltonians~\cite{murthy_srednicki_19a, nieva2023} in general does not produce the same structure in the Hilbert space as that introduced by the $U(1)$ symmetry. In the presence of latter, quantum states can be decomposed using direct sums of tensor products, which is not possible (in general) in the presence of the former. In the context of $SU(2)$ symmetry, in Ref.~\cite{patil2023} it was argued that ``interferences'' not accounted for in decompositions involving direct sums of tensor products can introduce $O(1)$ corrections. This motivates us to conjecture that the $O(1)$ correction in quantum-chaotic local Hamiltonians is not universal. To explore the validity of this conjecture, we plan to study the average entanglement entropy over eigenstates of different kinds of random matrices (SYK models being specific examples).

When considering fixed nonvanishing fractions of midspectrum Hamiltonian eigenstates in the average, $\nu\in(0,1]$, we found that the average eigenstate entanglement entropy differs by a $\nu$-dependent $O(1)$ correction from the prediction for the average over random states with increasing system size. For this case, we provided a simple expression for the $O(1)$ deviation expected from the maximal result at $\nu=0^+$ as a function of $\nu$. This phenomenological expression was obtained using the concave functional form of the eigenstate entanglement entropy, which exhibits a maximum in the middle of the spectrum of local quantum-chaotic Hamiltonians, and the known Gaussian form of the density of states in such models. Our analytical expression provides an excellent description of the numerical results for the average entanglement entropy, after numerically computing one parameter ($\bar S_A^{(\Lambda)}$) and fitting a second one ($c$) using the numerical results for the eigenstate entanglement entropy vs the eigenenergies. 

Our numerical results for a nonvanishing $\nu$ also indicate that the average entanglement entropy in eigenstates of the XYZ model differs from Huang's prediction (the average over eigenstates is greater) in the thermodynamic limit. Huang's $O(1)$ correction in the $\nu=1$ limit is identical to the $O(1)$ term derived in Ref.~\cite{bianchi_hackl_22} for the average over random pure states with fixed particle number (fixed magnetization in the spin language) when averaging over all particle-number sectors with a weight that is determined by the number of states in each sector. An interesting question to be explored in the future is what the $O(1)$ corrections at $\nu=0^+$ and nonvanishing $\nu$ can tell us about the Hamiltonian. For the (integrable) XY model in a transverse field, in Refs.~\cite{vidmar_hackl_18,hackl_vidmar_19} is was shown that there is an $O(1)$ correction at the critical line ($h = J$) for $f>0$, and no $O(1)$ correction otherwise, so that such a correction can be used to identify the critical line. 

Finally, we should mention that while the focus of this work was the nature of the $O(1)$ correction at $f=1/2$, we have also studied what happens when $f\neq1/2$. We briefly discuss those results in Appendix~\ref{sec:fneq05}. Our main finding away from $f=1/2$ is that, for a fixed number of midspectrum Hamiltonian eigenstates in the average ($\Lambda=100$), the average eigenstate entanglement entropy also appears to differ by a small $O(1)$ correction from the prediction for the average over random states in the thermodynamic limit. The difference obtained in our numerical calculations is clearly smaller than the one for $f=1/2$, and appears to decrease with decreasing the value of $f$. This parallels the behavior of the $O(1)$ mean-field term in Eq.~\eqref{eq:mfo1}. Determining the functional form of the $O(1)$ correction in model Hamiltonians as a function of $f$ is something that deserves a future investigation.

\acknowledgements
We acknowledge support from the Polish National Agency for Academic Exchange (NAWA) Grant No.~PPI/APM/2019/1/00085 (M.K.), the Slovenian Research Agency (ARRS), Research core fundings Grants No.~P1-0044 (R.S.~and L.V.) and No.~N1-0273 (L.V.), and the United States National Science Foundation (NSF) Grant No.~PHY-2012145 (M.R.). We acknowledge discussions with M. Haque. 

\appendix

\section{Maximally chaotic regime} \label{app::param_scan}

In Sec.~\ref{sec:chaos}, we described how we locate the \textit{maximally chaotic regime} and reported results for exemplary sets of parameters considered. Here we report results for the full set of parameters that we explored in some detail for the XYZ and the XXZ models. 

The XYZ model studied in this work has a large parameter space, with six independent parameters ($J_2,\, \eta,\, \Delta_1,\, \Delta_2,\, h^z,\,$ and $h_x$) after setting $J_1=1$ to be our energy scale [see Eq.~(\ref{model:eq1})]. We select $\eta = 0.5$, in the middle between the Ising point ($\eta=1$) and the XXZ point ($\eta=0$). In our early broader exploration of the space spanned by the other five parameters, we noticed that the results do not depend strongly on the strength of the transverse field $h^x$ unless it is made too large, which results in a departure from quantum chaotic behavior. Moreover, we found that when $\Delta_1$ and $\Delta_2$ are both $\sim$0.3, there is a wide range of values of $J_2$ and $h^z$ for which the numerical results for the various quantum chaos indicators considered here are close to the RMT predictions. This motivated us to set $h^x=0.3$, $\Delta_1=0.3$, and $\Delta_2 = 0.3$.

\begin{figure}[!t]
\includegraphics[width=0.85\columnwidth]{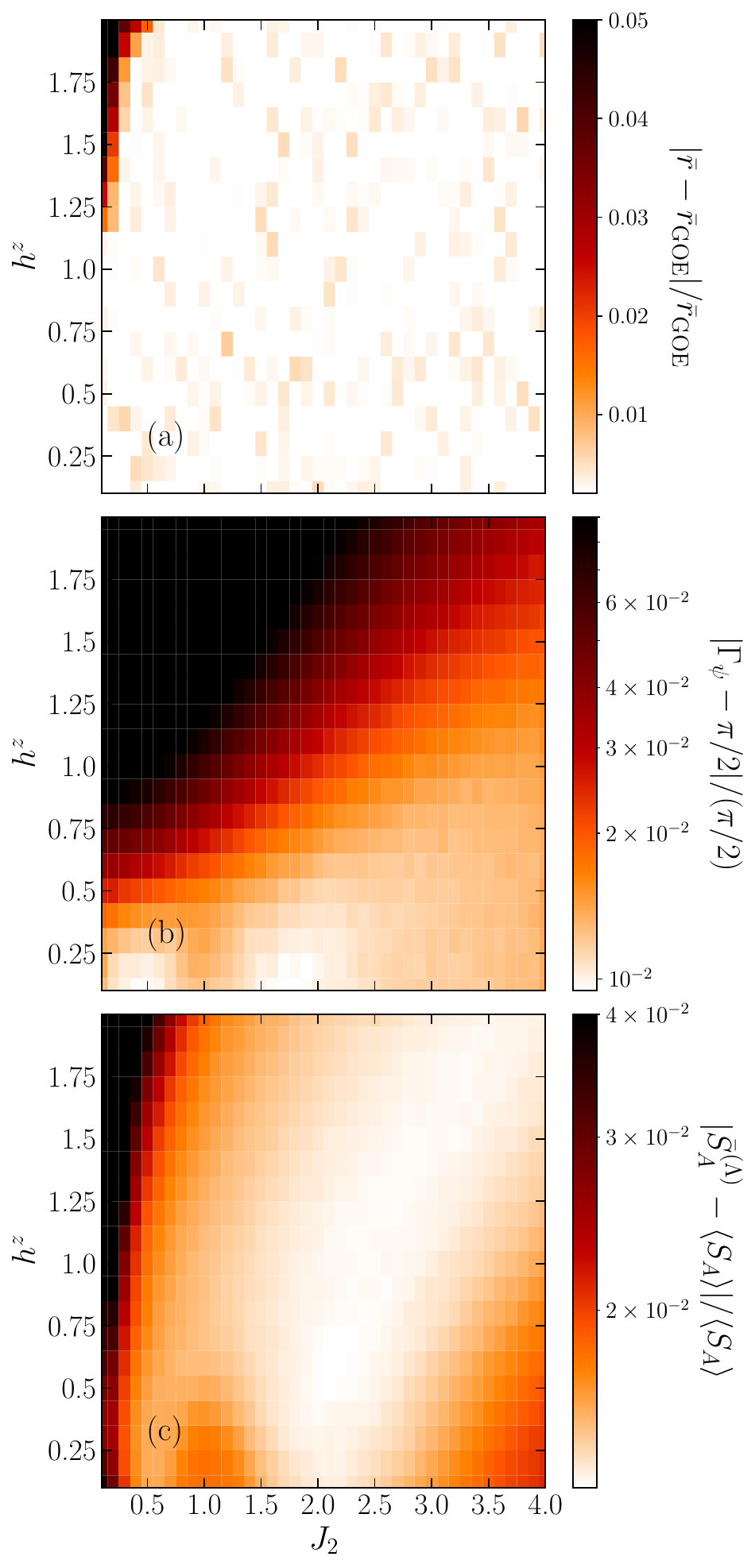}
\vspace{-0.2cm}
\caption{Normalized differences (a) $|\bar r - \bar r _{\rm GOE}|/\bar{r}_{\rm GOE}$, (b) $|\Gamma _\psi-\pi /2|/(\pi/2)$, and (c) $|\bar S_A ^{(\Lambda)} - \langle S _A \rangle|/\langle S _A \rangle$ for the ($J_2,\,h^z$) parameter space explored in the XYZ model, in chains with $L=18$ and PBCs. The differences are computed with respect to (a) the GOE prediction $\bar r _{\rm GOE}$, (b) the $\pi/2$ result for random states, and (c) Page's result $\langle S_A \rangle$ in Eq.~(\ref{eq:page_finite}).}
\label{app:param_scan:xyz}
\end{figure}

\begin{figure}[!t]
\includegraphics[width=0.85\columnwidth]{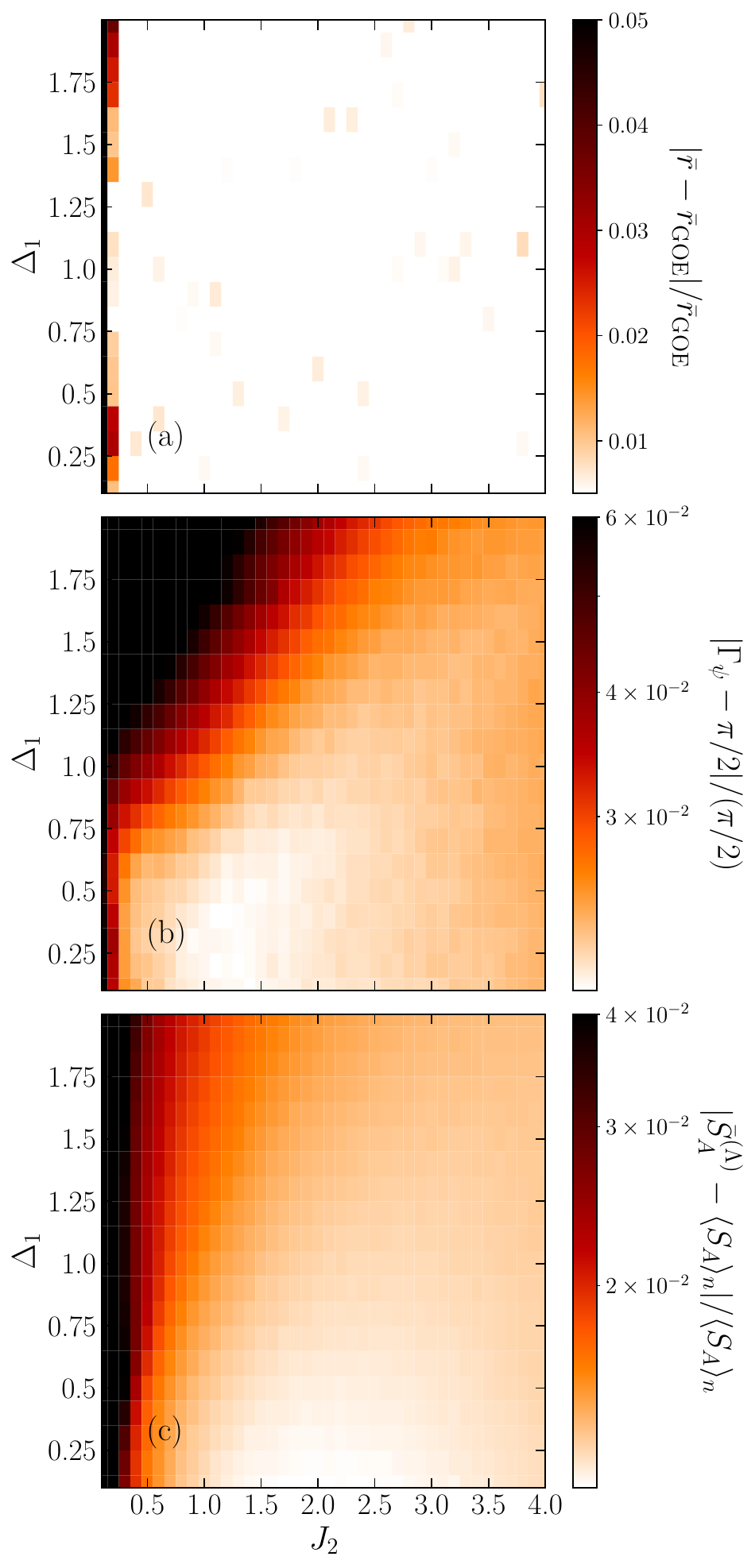}
\vspace{-0.2cm}
\caption{Normalized differences: (a) $|\bar r - \bar r _{\rm GOE}|/\bar{r}_{\rm GOE}$, (b) $|\Gamma _\psi-\pi /2|/(\pi/2)$, and (c) $|\bar S_A ^{(\Lambda)} - \langle S _A \rangle_n|/\langle S _A \rangle_n$ for the ($J_2,\,\Delta_1$) parameter space explored in the XXZ model, in chains with $L=20$ and PBCs. The differences are computed with respect to: (a) the GOE prediction $\bar r _{\rm GOE}$, (b) the $\pi/2$ result for random states, and (c) the result $\langle S_A \rangle_n$ in Eq.~(\ref{eq:page_n}).}
\label{app:param_scan:xxz}
\end{figure}

For XYZ chains with $L=18$ and PBCs, in Fig.~\ref{app:param_scan:xyz} we show density plots of the normalized differences between the numerical results and the RMT predictions for our quantum chaos indicators in the ($J_2,\,h^z$) plane. Figure~\ref{app:param_scan:xyz}(a) shows results for the average gap ratio [supplementing the results in Fig.~\ref{fig:levelspac}(a)], Fig.~\ref{app:param_scan:xyz}(b) shows results for our indicator of how close to Gaussian the distribution of eigenstate coefficients is [supplementing the results in the inset in Fig.~\ref{fig:coefficients}(a)], and Fig.~\ref{app:param_scan:xyz}(c) shows results for the average entanglement entropy [supplementing the results in Fig.~\ref{fig:aveentall}(a)]. In the latter case, the average entanglement entropy $\bar S_A^{(\Lambda)}$ is calculated using $\Lambda=100$ states in the middle of the spectrum from each symmetry block.

The results for all the quantum chaos indicators reported in Fig.~\ref{app:param_scan:xyz} consistently show that the largest deviations from the RMT predictions occur in the regime of small $J_2$ and large $h^z$. On the other hand, for identifying the maximally chaotic regime, the results reported in Figs.~\ref{app:param_scan:xyz}(b) and~\ref{app:param_scan:xyz}(c) are the most useful ones. We find the lowest normalized differences there to occur for $J_2\sim 2.0$ and $h^z<0.5$ (for the values of $h^z$ reported). This motivated us to selected $J_2 = 2.0$ and $h^z = 0.2$ for the scaling analysis of the average entanglement entropy discussed in Sec.~\ref{sec:entro}.  

The XXZ model studied in this work has three independent parameters ($J_2,\, \Delta_1,$ and $\Delta_2$) after setting $J_1=1$ to be our energy scale [see Eq.~(\ref{model:eq2})]. We set $\Delta_2 = 0.3$, as for the XYZ model, because it results in a wide range of values of $J_2$ and $\Delta_1$ for which the numerical results for the various quantum chaos indicators considered are close to the RMT predictions. For XXZ chains with $L=20$ and PBCs, in Fig.~\ref{app:param_scan:xxz} we show density plots of the normalized differences between the numerical results and the RMT predictions for our quantum chaos indicators in the ($J_2,\,\Delta_1$) plane. Those results parallel the ones in Fig.~\ref{app:param_scan:xyz} for the XYZ chains, and supplement the results reported in Fig.~\ref{fig:levelspac}(b), in the inset in Fig.~\ref{fig:coefficients}(b), and in Fig.~\ref{fig:aveentall}(b). For small values of $\Delta_1$ and intermediate values of $J_2\sim 2$, there is less structure in Figs.~\ref{app:param_scan:xxz}(b) and~\ref{app:param_scan:xxz}(c) than for small values of $h^z$ and intermediate values of $J_2\sim 2$ in Figs.~\ref{app:param_scan:xyz}(b) and~\ref{app:param_scan:xyz}(c). From the wider range of parameters that give results in the XXZ chain that are similarly close to the RMT predictions, we selected $J_2=2.0$ and $\Delta _1=0.2$ for the scaling analysis of the average entanglement entropy discussed in Sec.~\ref{sec:entro}.  

\section{Distribution of eigenstate coefficients} \label{app::dist}

In Sec.~\ref{sec:coeff}, we reported results for the distributions of the absolute values of the scaled real and imaginary parts of the eigenstate coefficients in the computational basis, which we denoted as $z$. Those distributions were compared to the Gaussian distribution in Eq.~(\ref{coeff:eq1}). The latter distribution was derived assuming that the scaled real and imaginary parts $x$ of the eigenstate coefficients are Gaussian with variance 1,
\begin{eqnarray}
    \bar P_X(x) = \frac{1}{\sqrt{2\pi}}e^{-x^2/2}\;.
\end{eqnarray}

To derive $P_Z(z)$, we note that $z = g(x) = |x|$. In general, if $g(x)$ is an invertible function, then one can calculate the probability distribution of $z$ using the formula
\begin{equation}\label{app:coeff:eq1}
    \bar P_Z(z=g(x))= \bar P_X(g^{-1}(z))\abs{\frac{dg^{-1}(z)}{dz}}\;,
\end{equation}
and, for piecewise invertible functions, given the set for $x=h(z)$
\begin{equation*}
    {h_i(z):\ \ \exists_{x_i^1,x_i^2}\ h^{-1}_i(x)=g(x) \text{ for } x\in(x_i^1,x_i^2)}\;,
\end{equation*}
Eq.~\eqref{app:coeff:eq1} can be generalized as
\begin{equation}\label{app:coeff:eq2}
    \bar P_Z(z=g(x))=\sum_i \bar P_X(h_i(z))\abs{\frac{dh_i(z)}{dz}}\;.
\end{equation}
Hence, the distribution of $z$ is given by
\begin{equation}
    \bar P_Z(z) = \bar P_X(z) + \bar P_X(-z) = \frac{1}{\sqrt{2\pi}}e^{-z^2/2} + \frac{1}{\sqrt{2\pi}}e^{-(-z)^2/2},
\end{equation}
which yields the result in Eq.~\eqref{coeff:eq1}.

\begin{figure}[!t]
\includegraphics[width=0.98\linewidth]{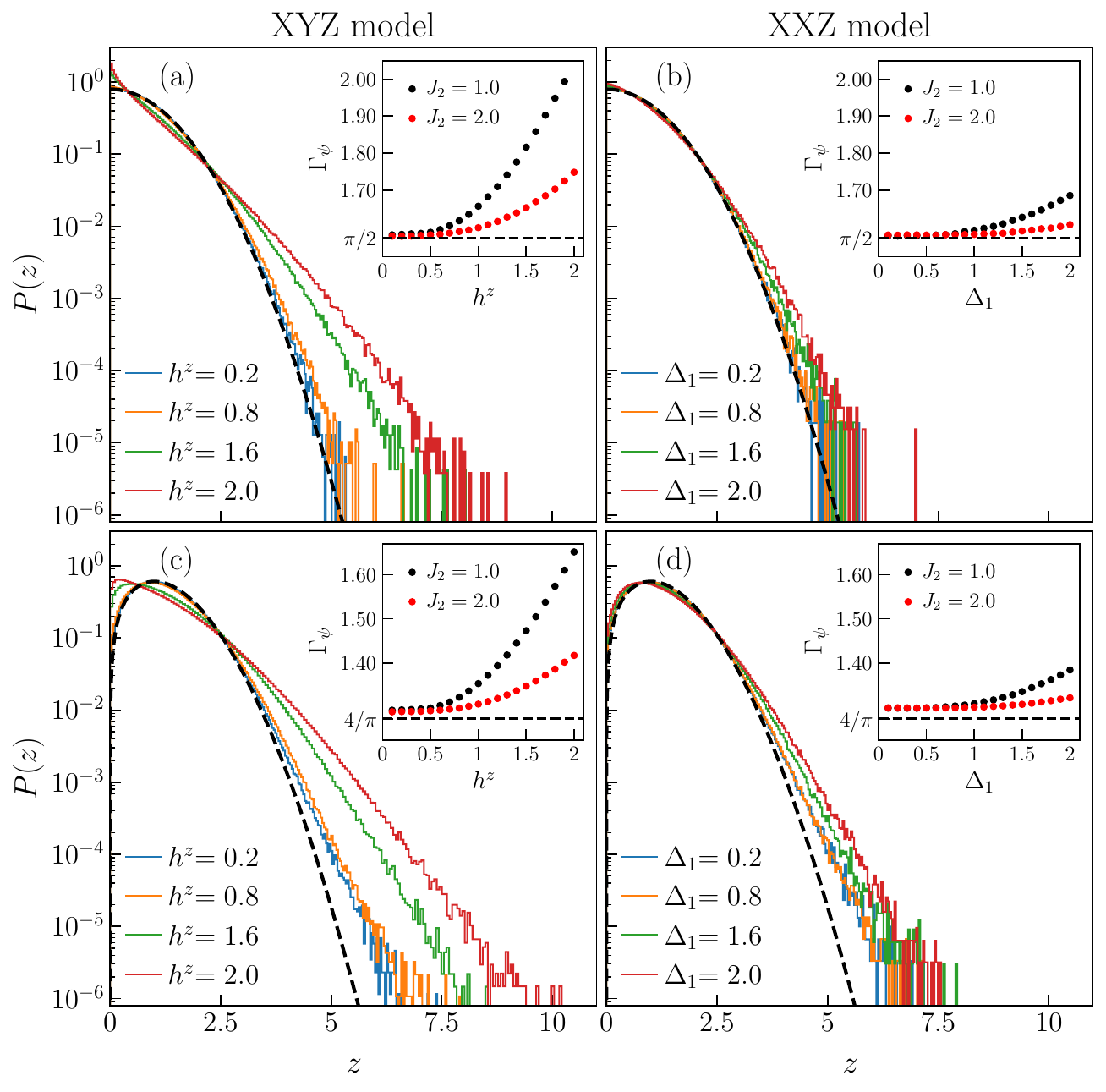}
\vspace{-0.2cm}
\caption{Distribution of the absolute values of eigenstate coefficients collected from 100 midspectrum eigenstates of each symmetry block from the: [(a) and (b)] $k=0,\,\pi$ (``real'') and [(c) and (d)] all other ($k\neq 0,\,\pi$; ``complex'') quasimomentum sectors. The results were obtained taking $J_2 = 1.0$, for [(a) and (c)] the XYZ model ($L=18$) and [(b) and (d)] the XXZ model ($L=20$). The dashed lines show $P(z)$ from Eq.~\eqref{coeff:eq1} in (a) and (b) and from Eq.~\eqref{app:coeff:eq7} in (c) and (d). The insets show the ratio $\Gamma_\psi$ as a function of $h^z$ in (a) and (c) and of $\Delta_1$ in (b) and (d). The horizontal dashed lines are the predictions for random states [see discussion in Sec.~\ref{sec:coeff} and Eq.~\eqref{eq:gamma_psi_2}].}
\label{app:coeff:dist}
\end{figure}

Instead of studying the distributions of the absolute values $z$ of the scaled real and imaginary parts of the eigenstate coefficients in the computational basis, which allowed us to treat all the quasimomentum sectors on an equal footing, one can separately study the distributions of the absolute values $z$ of the scaled real eigenstate coefficients in the $k=0, \, \pi$ (``real'') quasimomentum sectors, and separately of the absolute values $\tilde z$ of the scaled complex eigenstate coefficients in all the other ($k\neq 0, \, \pi$; ``complex'') quasimomentum sectors.

Let us derive the PDF of the absolute value $\tilde z$ of the complex coefficients in the $k\neq 0, \, \pi$ sectors. We define $\tilde z=|x+iy|$, where $x$ is the real part of the coefficients [$\Re(c_m^\alpha)$] and $y$ is the imaginary part of the coefficients [$\Im(c_m^\alpha)$]. As a first step, let us find the PDF of $w = \tilde z^2= x^2 + y^2$. Under the assumption that both $x$ and $y$ are normally distributed with zero mean and variance 1, the PDF of $w$ is the chi-squared distribution of degree 2,
\begin{equation}
    \bar P_{W} (w) = \frac{1}{2}e^{-\frac{w}{2}}\;.
\end{equation}
Then, using Eq.~(\ref{app:coeff:eq1}), we obtain that
\begin{equation}\label{app:coeff:eq7}
   \bar P_{\tilde Z}(\tilde z)=\tilde z\, e^{-\frac{{\tilde z}^2}{2}}.
\end{equation}

Next, let us find the ratio $\Gamma_\psi$ defined in Eq.~\eqref{coeff:eq2} for the distribution in Eq.~\eqref{app:coeff:eq7}. The first moment of $\tilde z$ is
\begin{equation}
 \expval{\tilde z}=\int_0^\infty d\tilde z\, \tilde z \,P_{\tilde Z}(\tilde z)=\int_0^\infty d \tilde z \,{\tilde z}^2\,e^{-\frac{{\tilde z}^2}{2}}=\sqrt{\frac{\pi}{2}} \;,
\end{equation}
and the second moment is
\begin{equation}
        \expval{{\tilde z}^2}=\int_0^\infty d {\tilde z}\, {\tilde z}^2\, {\tilde z}\, e^{-\frac{{\tilde z}^2}{2}}=2 \;,
\end{equation}
so that the ratio in Eq.~\eqref{coeff:eq2} is
\begin{equation} \label{eq:gamma_psi_2}
    \Gamma_\psi=\frac{\expval{{\tilde z}^2}}{\expval{{\tilde z}}^2}=\frac{4}{\pi}\;.
\end{equation}

In Figs.~\ref{app:coeff:dist}(a) and~\ref{app:coeff:dist}(c) [Figs.~\ref{app:coeff:dist}(b) and~\ref{app:coeff:dist}(d)], we show results obtained for the distributions of the scaled (by the corresponding standard deviation) absolute values of eigenstate coefficients for the XYZ model in chains with $L=18$ [the XXZ model in chains with $L=20$] and PBCs. The top panels [Figs.~\ref{app:coeff:dist}(a) and~\ref{app:coeff:dist}(b)] show results for the $k=0,\, \pi$ quasimomentum sectors compared to the PDF (dashed line) in Eq.~\eqref{coeff:eq1}, while the bottom panels [Figs.~\ref{app:coeff:dist}(c) and~\ref{app:coeff:dist}(d)] show results for the $k\neq 0,\, \pi$ quasimomentum sectors compared to the PDF (dashed line) in Eq.~\eqref{app:coeff:eq7}. Results for $\Gamma_\psi$ are shown in the insets, in which the horizontal lines mark the prediction for the PDF of Eq.~(\ref{coeff:eq1}) in Figs.~\ref{app:coeff:dist}(a) and~\ref{app:coeff:dist}(b), and the prediction of Eq.~(\ref{eq:gamma_psi_2}) in Figs.~\ref{app:coeff:dist}(c) and~\ref{app:coeff:dist}(d). These results can be seen as being complementary to the ones in Fig.~\ref{fig:coefficients}, where we reported results for the distribution of the absolute values of the scaled real and imaginary parts of the eigenstate coefficients from all symmetry blocks (all values of $k$) bundled together.

\begin{figure}[!t]
\includegraphics[width=0.85\linewidth]{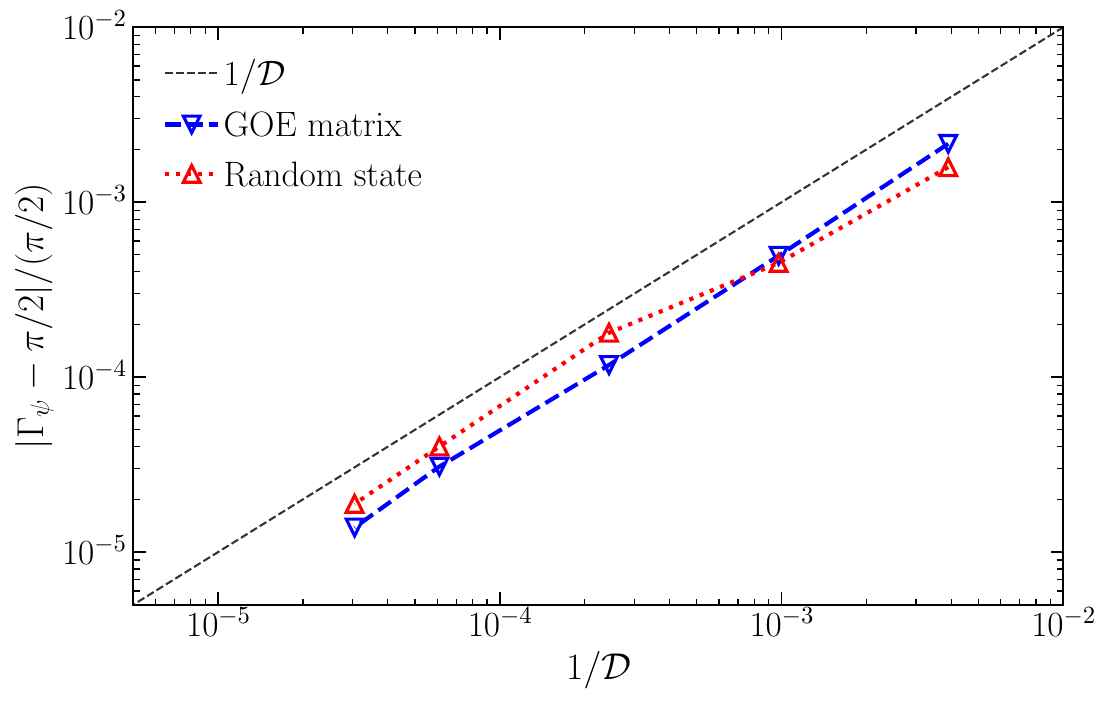}
\vspace{-0.2cm}
\caption{Scaling of $|\Gamma_\psi-\pi /2|/(\pi/2)$ for midspectrum eigenstates of matrices that belong to the GOE and for random pure states whose coefficients are sampled from a normal distribution. The normalized differences are plotted as functions of $1/\mathcal{D}$, where $\mathcal{D}$ is the dimension of the Hilbert space. Taking $\mathcal{D}=2^L$, the dimensions shown would correspond to chains with $L=8,\,10,\,12,\,14,$ and 15. We compute $\Gamma_\psi$ by (i) diagonalizing 100 GOE matrices and taking $100$ midspectrum eigenstates from each of them, and (ii) by generating $10^4$ realizations of the random pure states. The black dotted line indicates $\sim\mathcal{D}^{-1}$ behavior.}
\label{app:coeff:random}
\end{figure} 

The results in Fig.~\ref{app:coeff:dist} suggest that the convergence to the predictions for random states is faster in the $k=0,\, \pi$ sectors, compare the agreement with the theoretical predictions seen in Figs.~\ref{app:coeff:dist}(a) and~\ref{app:coeff:dist}(b) vs in Figs.~\ref{app:coeff:dist}(c) and~\ref{app:coeff:dist}(d), even though for any given value of $L$ the ``real'' sectors have 1/2 of the number of states in the other sectors (due to the presence of reflection symmetry). In the maximally chaotic regime, the PDFs in the XYZ model are the ones that we find to be closest to the theoretical predictions. Notice that, for the parameters shown, the set selected for the scaling in Sec.~\ref{sec:entro} is the one that gives the closest results to the theoretical predictions.

We also note that for the results shown in the insets in Figs.~\ref{fig:coefficients} and~\ref{app:coeff:dist}, even in the maximally chaotic regime of both models, $\Gamma_\psi$ is slightly larger than the theoretical prediction for random states. In Appendix~\ref{sec:app_scaling}, we report the scaling of the normalized difference $|\Gamma_\psi-\pi/2|/(\pi/2)$ for the XYZ model in chains with both periodic and open boundary conditions. They show that, as expected, the normalized differences decrease with increasing system size. In Fig.~\ref{app:coeff:random}, we show how $|\Gamma _\psi-\pi /2|/(\pi/2)$ depends on the inverse Hilbert space dimension $1/\mathcal{D}$ for midspectrum eigenstates of matrices that belong to the GOE and for random pure states whose coefficients are sampled from a normal distribution. In those two cases, the deviations from the predicted value of $\pi/2$ in the thermodynamic limit are much smaller than for Hamiltonian eigenstates and decrease as $1/\mathcal{D}$.

\section{Scaling of the entanglement entropy} \label{sec:app_scaling}

\begin{figure}[!b]
\includegraphics[width=0.98\linewidth]{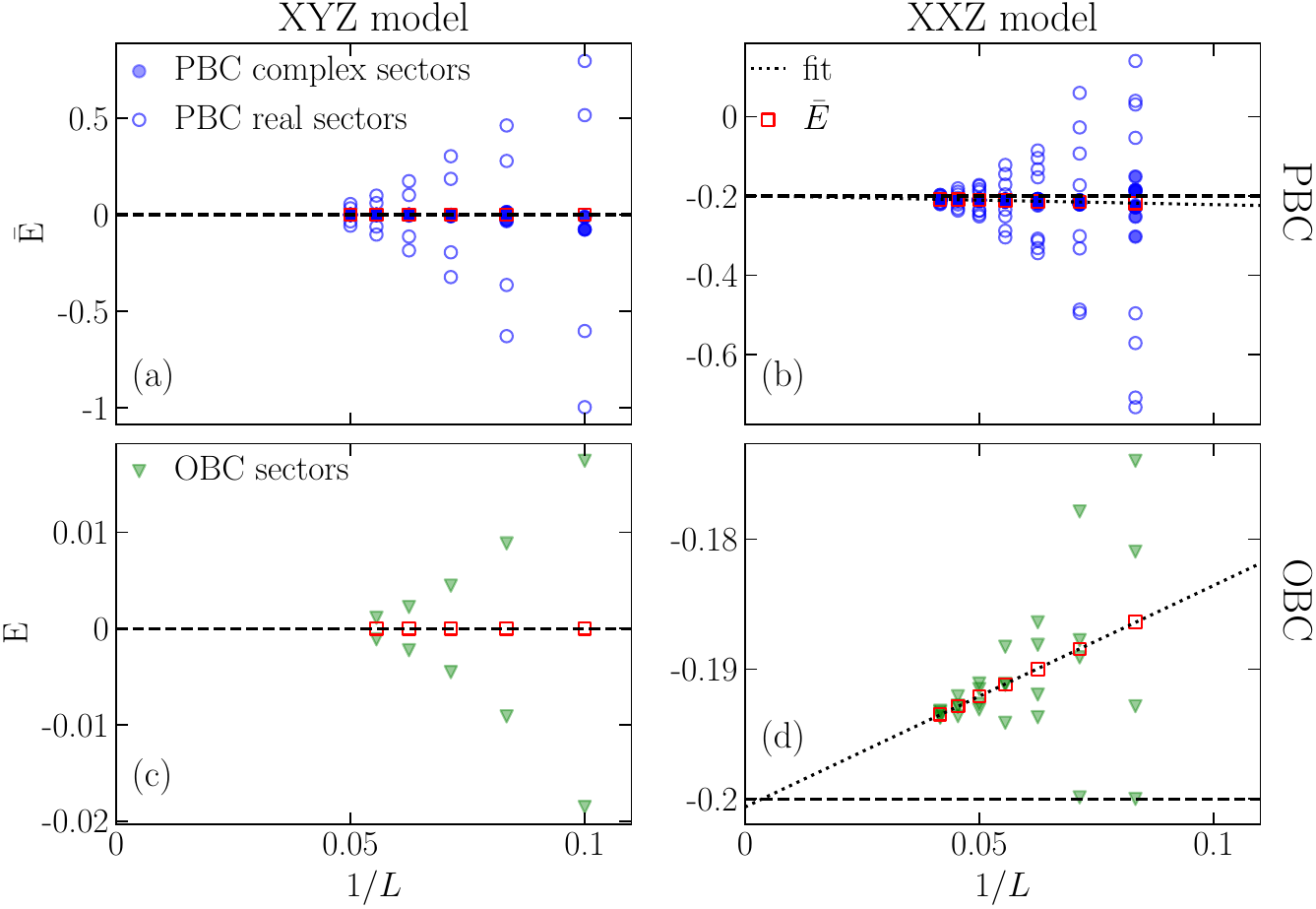}
\vspace{-0.2cm}
\caption{Scaling of the mean energy $\bar {\mathrm E}$ in all symmetry blocks vs $1/L$ for [(a) and (c)] XYZ and [(b) and (d)] XXZ chains with [(a) and (b)] periodic and [(c) and (d)] open boundary conditions, for the same Hamiltonian parameters used in Sec.~\ref{sec:entro}. The empty red squares show the average energy over the entire Hilbert space $\bar E$, and the horizontal dashed lines mark the theoretical prediction for $\bar E$ in the thermodynamic limit: [(a) and (c)] $\bar E^\text{XYZ}_\infty=0$ and [(b) and (d)] $\bar E^\text{XXZ}_\infty=-\frac{1}{4}\qty(J_1\Delta_1+J_2\Delta_2)=-0.2$. In (b) and (d), the dotted lines show fits of the numerical results for $\bar E$ in finite XXZ chains to the function $c_1+c_2/L$, which returned (b) $c_1=-0.199$ and $c_2=-0.227$, and (d) $c_1=-0.201$ and $c_2=0.171$. In (a) and (b), we use empty (filled) symbols to report $\bar {\mathrm E}$ in the $k=0,\,\pi$ ($k\neq 0,\,\pi$) quasimomentum blocks.}
\label{fig:app:escaling}
\end{figure} 

As mentioned in the main text, when computing the eigenstate entanglement entropy averages we define the ``midspectrum eigenstates'' for each symmetry block separately, i.e., they are the eigenstates whose eigenenergies are closest to the mean energy $\bar {\mathrm E}$ in each symmetry block. This, as opposed to using the mean energy $\bar E={\text{Tr}}(\hat{H})/{\mathcal{D}}$ in the entire Hilbert space, is done as a way to reduce finite-size effects in the smallest chains considered. In the thermodynamic limit, $\bar E$ for the XYZ model is $\bar E^\text{XYZ}_\infty=0$, while for the XXZ model, see Eq.~(\ref{model:eq2}), it is $\bar E^\text{XXZ}_\infty=-\frac{1}{4}\qty(J_1\Delta_1+J_2\Delta_2)$.

\begin{figure}[!t]
\includegraphics[width=0.98\linewidth]{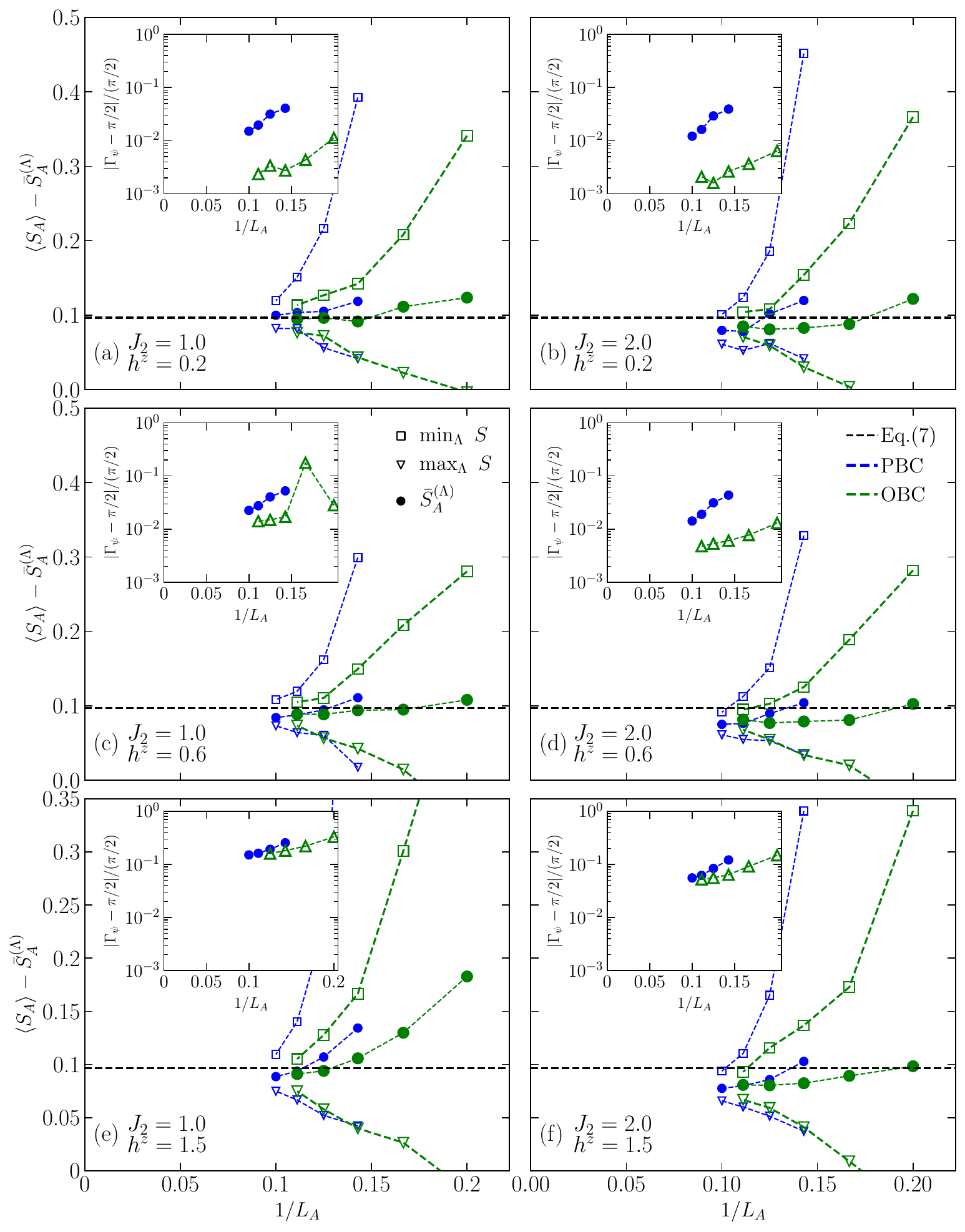}
\vspace{-0.2cm}
\caption{Finite-size scaling analysis of the average entanglement entropy of $\Lambda=100$ midspectrum eigenstates in XYZ chains with various sets of parameters. The deviations of the average eigenstate entanglement entropy $\bar S_A^{(\Lambda)}$ (filled symbols), and the outlier eigenstate entanglement entropies (open symbols), from the exact result for random pure states $\langle S_A\rangle$ [Eq.~\eqref{eq:page_finite}] are plotted vs the inverse subsystem size $L_A=L/2$. The insets show $|\Gamma_\psi-\pi /2|/(\pi/2)$ for the same parameters as in the main panels vs $1/L_A$. Results are shown for chains with both periodic and open boundary conditions. The horizontal dashed lines show the predictions of Eq.~\eqref{eq:nu}. The results in (b) are the same as in Fig.~\ref{fig:entropyscalXYZ}(a).}
\label{app:param_scan:entxyz}
\end{figure}

\begin{figure}[!t]
\includegraphics[width=0.98\linewidth]{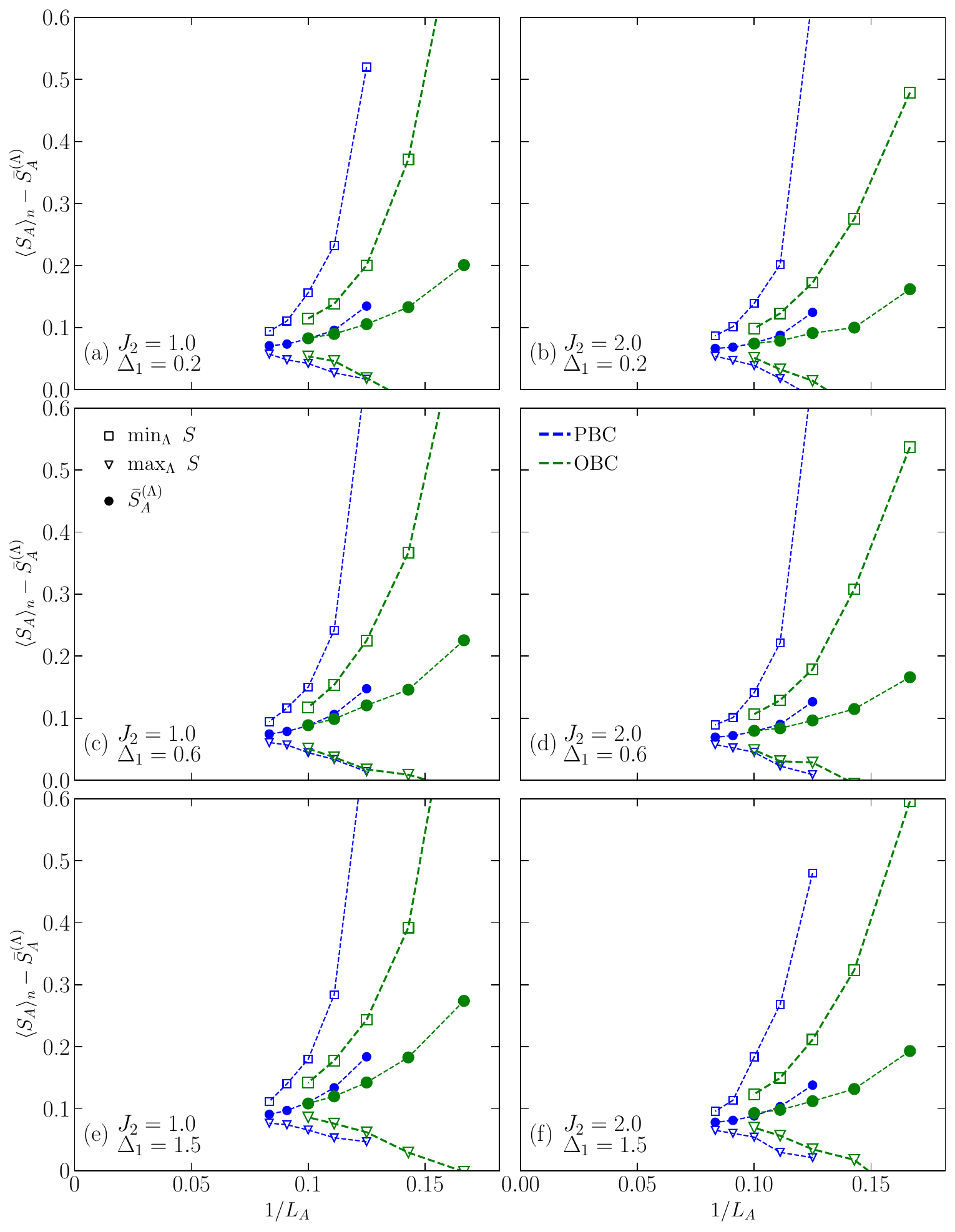}
\vspace{-0.2cm}
\caption{Finite-size scaling analysis of the average entanglement entropy of $\Lambda=100$ midspectrum eigenstates in XXZ chains with various sets of parameters. The deviations of the average eigenstate entanglement entropy $\bar S_A^{(\Lambda)}$ (filled symbols), and the outlier eigenstate entanglement entropies (open symbols), from the exact result for random pure states $\langle S_A\rangle_n$ [Eq.~\eqref{eq:page_n}] are plotted vs the inverse subsystem size $L_A=L/2$. Results are shown for chains with both periodic and open boundary conditions. The results in (b) are the same as in Fig.~\ref{fig:entropyscalXXZ}(a).}
\label{app:param_scan:entxxz}
\end{figure}

In Fig.~\ref{fig:app:escaling}, we show the scaling of the mean energy $\bar {\mathrm E}$ in all symmetry blocks vs $1/L$. The plots make apparent that, with increasing system size, $\bar {\mathrm E}$ in all symmetry blocks rapidly collapses onto $\bar E$ (plotted as empty red squares), both for the XYZ [Fig.~\ref{fig:app:escaling}(a)] and the XXZ [Fig.~\ref{fig:app:escaling}(b)] models in chains with periodic and open boundary conditions. For the XXZ chains, one can further see that $\bar E$ approaches $\bar E^\text{XXZ}_\infty$ polynomially with increasing system size, as expected for canonical ensemble calculations~\cite{iyer_15}. Linear fits of the results for $\bar E$ in finite systems return a thermodynamic limit result that is almost in perfect agreement with the theoretical prediction. Finally, we note that in Figs.~\ref{fig:app:escaling}(a) and~\ref{fig:app:escaling}(b), the results for $\bar {\mathrm E}$ in the $k\neq 0,\,\pi$ quasimomentum blocks are much closer to $\bar E$ than those for $\bar {\mathrm E}$ in the $k= 0,\,\pi$ blocks.

Next, we report results for the scaling of the average entanglement entropy of $\Lambda=100$ midspectrum eigenstates in XYZ and XXZ chains with various sets of parameters. They show that the results reported in Sec.~\ref{sec:entro} are robust against changes in the Hamiltonian parameters, so long as the parameters are not taken to be too far from the maximally chaotic regime.

In Fig.~\ref{app:param_scan:entxyz}, we plot $\langle S _A\rangle - \bar S_A^{(\Lambda)}$ for the XYZ model for six pairs of values of $(J_2,\,h^z)$. The results in Fig.~\ref{app:param_scan:entxyz}(b) are the same as in Fig.~\ref{fig:entropyscalXYZ}(a). The theoretical results for the random pure states $\langle S _A\rangle$ are the ones predicted by Eq.~(\ref{eq:page_finite}). As in Fig.~\ref{fig:entropyscalXYZ}, we include the maximal and minimal outliers, which bound the eigenstate entanglement entropies that enter the averages. In Fig.~\ref{app:param_scan:entxyz}, $J_2$ increases from left to right, while $h^z$ increases from top to bottom, and the horizontal dashed lines mark the result from Eq.~\eqref{eq:nu}. The insets show the scaling of $|\Gamma_\psi-\pi /2|/(\pi/2)$ for the same parameters as in the main panels vs $1/L_A$. 

All the results in the main panels of Fig.~\ref{app:param_scan:entxyz} suggest that, with increasing system size, $\bar S_A^{(\Lambda)}$ approaches a value that is slightly smaller than the prediction of Eq.~(\ref{eq:page_finite}). The difference appears to be smaller than $0.1$ in all panels but in Fig.~\ref{app:param_scan:entxyz}(a). The results for $|\Gamma_\psi-\pi /2|/(\pi/2)$ in the insets depend more strongly on the parameters chosen, with small $h^z$ resulting in the smallest normalized differences, and are consistent with a polynomial decrease with $1/L_A=2/L$ as $L$ increases.

In Fig.~\ref{app:param_scan:entxxz}, we plot $\langle S _A\rangle_n - \bar S_A^{(\Lambda)}$ for the XXZ model for six pairs of values of $(J_2,\,\Delta_1)$. The results in Fig.~\ref{app:param_scan:entxxz}(b) are the same as in Fig.~\ref{fig:entropyscalXXZ}(a). The theoretical results for the random pure states $\langle S _A\rangle_n$ are the ones predicted by Eq.~(\ref{eq:page_n}). As in Fig.~\ref{fig:entropyscalXXZ}, we include the maximal and minimal outliers, which bound the eigenstate entanglement entropies that enter the averages. In Fig.~\ref{app:param_scan:entxxz}, $J_2$ increases from left to right, while $\Delta_1$ increases from top to bottom. All the results in Fig.~\ref{app:param_scan:entxxz} suggest that, with increasing system size, $\bar S_A^{(\Lambda)}$ approaches a value that is slightly smaller than the predictions of Eq.~(\ref{eq:page_finite}). The difference appears to be smaller than $0.1$ in all panels.

\section{XYZ eigenstates vs random states\\ at fixed zero magnetization} \label{sec:EfromN} 

\begin{figure}[!b]
\includegraphics[width=0.98\linewidth]{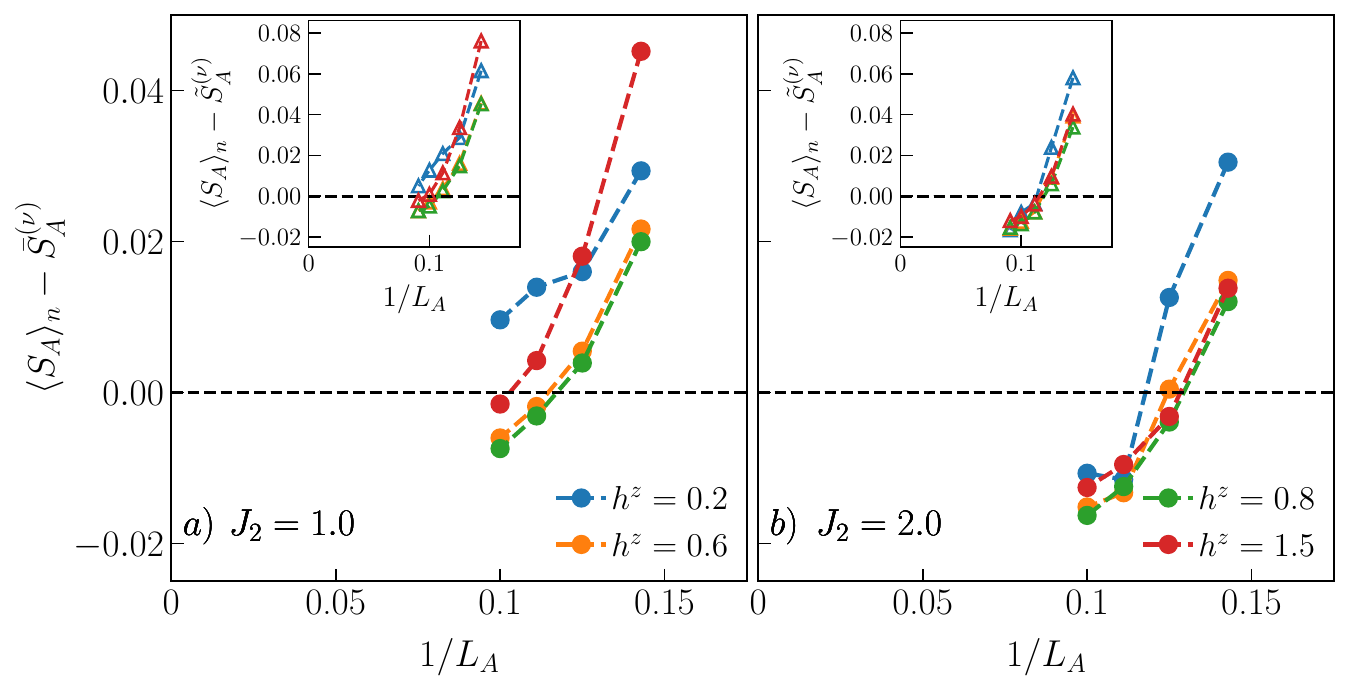}
\vspace{-0.2cm}
\caption{Additional finite-size scaling analysis of the average entanglement entropy of $\Lambda=100$ midspectrum eigenstates in XYZ chains with periodic boundary conditions. The deviations of the average eigenstate entanglement entropy $\bar S_A^{(\Lambda)}$ from the exact result for random pure states at fixed zero magnetization $\langle S_A\rangle_n$ [Eq.~\eqref{eq:page_n}] are plotted vs the inverse subsystem size $L_A=L/2$. The main panels show the numerical results for the averages over all symmetry sectors, while the insets show the average over ``real'' sectors ($k=0$ and $\pi$ sectors, for which we can diagonalize the largest chains). We report results for (a) $J_1=1.0$ and (b) $J_2=2.0$ for different values of $h^z$.}
\label{app:param_scan:entxyz_vs_U1}
\end{figure}

Since Huang's $O(1)$ correction at $\nu=0^+$~\cite{huang_21} is identical to that predicted for random pure states at zero magnetization~\cite{vidmar_rigol_17, bianchi_hackl_22}, in Fig.~\ref{app:param_scan:entxyz_vs_U1} we provide an additional finite-size scaling analysis of the average entanglement entropy of $\Lambda=100$ midspectrum eigenstates in XYZ chains with periodic boundary conditions when compared to the exact result for random pure states at fixed zero magnetization $\langle S_A\rangle_n$ [see Eq.~\eqref{eq:page_n}]. The differences are clearly smaller than when comparing to the exact result for random pure states $\langle S_A\rangle$ [see Eq.~\eqref{eq:page_finite}], but we find no indication that they will vanish in the thermodynamic limit. For the parameters considered, our numerical results suggest that ${\bar S}_A^{(\Lambda)}>\langle S_A\rangle_n$ for $L\rightarrow\infty$.

\section{Subsystem fractions $f\neq 1/2$} \label{sec:fneq05}

All the previous results for the average eigenstate entanglement entropy were computed for a subsystem fraction $f=L_A/L = 1/2$. Here we report results obtained for smaller subsystem fractions. For the XYZ model, for which there is no restriction on the value of the magnetization and hence for which we can study chains with even and odd numbers of sites, we consider $f=1/3$ and $f=1/4$. For the XXZ model, for which we restrict our study to the zero total magnetization sector and hence only study chains with an even number of sites, we only consider $f=1/4$.

\begin{figure}[!b]
\includegraphics[width=0.99\linewidth]{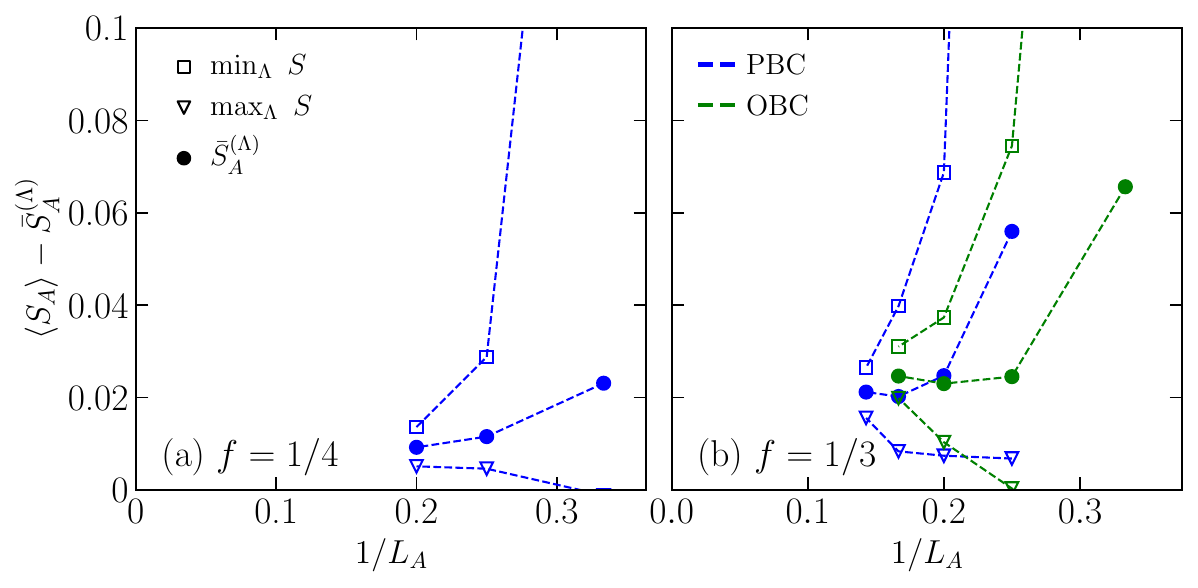}
\vspace{-0.2cm}
\caption{Finite-size scaling analysis of the average entanglement entropy of $\Lambda=100$ midspectrum eigenstates in XYZ chains with $J_2=2.0$ and $h^z=0.2$. The deviations of the average eigenstate entanglement entropy $\bar S_A^{(\Lambda)}$ (filled symbols), and the outlier eigenstate entanglement entropies (open symbols), from the exact result for random pure states $\langle S_A\rangle$ [Eq.~\eqref{eq:page_finite}] are plotted vs the inverse subsystem size $L_A=fL$. The subsystem fractions are: (a) $f=1/4$, for which we only show results for PBCs because of the limited sizes accessible for OBCs, and (b) $f=1/3$, for which we show results for both PBCs and OBCs.}
\label{app:frac_xyz}
\end{figure}

Figure~\ref{app:frac_xyz} shows results for the XYZ model in the maximally chaotic regime when $J_2=2.0$ and $h^z=0.2$ (same parameters as in Sec.~\ref{sec:xyz}), for $f=1/4$ [Fig.~\ref{app:frac_xyz}(a)] and $f=1/3$ [Fig.~\ref{app:frac_xyz}(b)]. We again find indications of an $O(1)$ difference in the thermodynamic limit, but with a value that appears to depend on $f$. Notice that the deviation from the results for random pure states $\langle S_A \rangle$ is smaller at $f=1/4$ [Fig.~\ref{app:frac_xyz}(a)], than at $f=1/3$ [Fig.~\ref{app:frac_xyz}(b)]. In both cases the differences are much smaller than at $f=1/2$. This suggests that if the difference remains nonzero in the thermodynamic limit, then the $O(1)$ contribution is likely to decrease with decreasing $f$.

\begin{figure}[!t]
\includegraphics[width=0.75\linewidth]{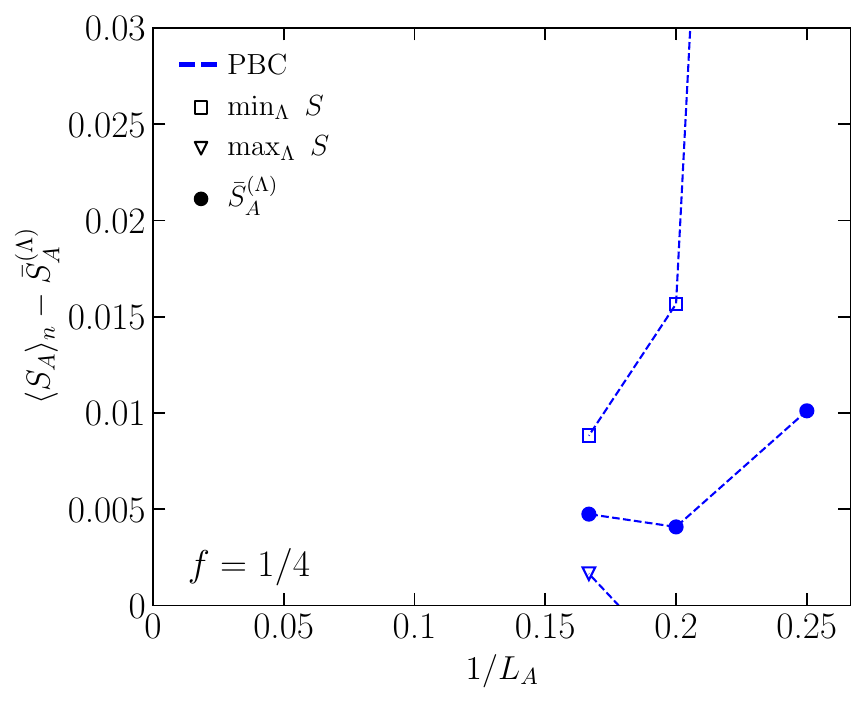}
\vspace{-0.2cm}
\caption{Finite-size scaling analysis of the average entanglement entropy of $\Lambda=100$ midspectrum eigenstates in XXZ chains with $J_2=2.0$ and $\Delta_1=0.2$ at $f=1/4$. The deviations of the average eigenstate entanglement entropy $\bar S_A^{(\Lambda)}$ (filled symbols), and the outlier eigenstate entanglement entropies (open symbols), from the exact result for random pure states $\langle S_A\rangle_n$ [Eq.~\eqref{eq:page_n}] are plotted vs the inverse subsystem size $L_A=L/4$. We only show results for PBCs because of the limited sizes accessible for OBCs.}
\label{app:frac_xxz}
\end{figure}

Figure~\ref{app:frac_xxz} shows results for the XXZ model in the maximally chaotic regime when $J_2=2.0$ and $\Delta_1=0.2$ (same parameters as in Sec.~\ref{sec:xxz}). Only three system sizes are accessible for PBCs at $f=1/4$, but the results are qualitatively and quantitatively similar to those for the XYZ model at $f=1/4$ in Fig.~\ref{app:frac_xyz}(a). Together with the results for the XYZ model at $f=1/3$ in Fig.~\ref{app:frac_xyz}(b), our numerical results suggest that the $O(1)$ difference, if nonzero in the thermodynamic limit, is much smaller for $f<1/3$ than at $f=1/2$.

\newpage

\bibliographystyle{biblev1}
\bibliography{references}

\end{document}